\documentclass[twocolumn]{aastex63}
\usepackage{amsmath,amstext}
\usepackage[T1]{fontenc}
\usepackage{apjfonts} 
\usepackage{soul}
\usepackage[figure,figure*]{hypcap}

\received{~--- xxx}
\revised{~--- xxx}
\accepted{~--- xxx}

\shorttitle{Compact star-forming galaxies: Blueberry}
\shortauthors{Paswan et al.}

\begin{document}

\title{\large SDSS-IV MaNGA: an observational evidence of density bounded region in a Lyman-$\alpha$ emitter}

\correspondingauthor{Abhishek Paswan}
\email{paswanabhishek@iucaa.in}

\author{Abhishek Paswan}
\affil{Inter-University Centre for Astronomy and Astrophysics, Ganeshkhind, Post Bag 4, Pune 411007, India}

\author{Kanak Saha}
\affiliation{Inter-University Centre for Astronomy and Astrophysics, Ganeshkhind, Post Bag 4, Pune 411007, India}

\author{Claus Leitherer}
\affiliation{Space Telescope Science Institute, Baltimore, MD 21218, USA}

\author{Daniel Schaerer}
\affiliation{Observatoire de Genève, Université de Genève, 51 Ch. des Maillettes, 1290, Versoix, Switzerland}

\begin{abstract}

Using Integral Field Unit (IFU) spectroscopy, we present here the spatially resolved morphologies of [SII]$\lambda$6717,6731/H$\alpha$ and [SII]$\lambda$6717,6731/[OIII]$\lambda$5007 emission line ratios for the first time in a Blueberry Lyman-$\alpha$ emitter (BBLAE) at z $\sim$ 0.047. Our derived morphologies show that the extreme starburst region of the BBLAE, populated by young ($\leqslant$ 10 Myr), massive Wolf-Rayet stars, is [SII]-deficient, while the rest of the galaxy is [SII]-enhanced. We infer that the extreme starburst region is density-bounded (i.e., optically thin to ionizing photons), and the rest of the galaxy is ionization-bounded $-$ indicating a Blister-type morphology. We find that the previously reported small escape fraction (10\%) of Ly$\alpha$ photons is from our identified density-bounded H{\sc ii} region of the BBLAE. This escape fraction is likely constrained by a porous dust distribution.  
\par
We further report a moderate correlation between [SII]-deficiency and inferred Lyman Continuum (LyC) escape fraction using a sample of confirmed LyC leakers studied in the literature, including the BBLAE studied here. The observed correlation also reveals its dependency on the stellar mass and gas-phase metallicity of the leaky galaxies. Finally, the future scope and implications of our work are discussed in detail.
\end{abstract}


\section{Introduction} 
\label{sec:intro}

In recent years, a lot of efforts have been made to identify the sources responsible for the re-ionization of the early Universe (z $\gtrsim$ 6). Although active galactic nuclei (AGN) are one of the sources of re-ionization, their contribution remains debated \citep{Madau2015,Hassan2018}. High redshift, low-mass compact starburst galaxies are now thought to be one of the significant contributors to re-ionization \citep{Ouchi2009,Mitra2013,Naidu2018,Kimm2019}. Besides the identification of ionizing sources, the specific physical processes that allow the escape of ionizing photons from these sources are also poorly understood. Several observational studies in the literature have confirmed that galaxies exhibiting leakage of ionizing photons often show a high value of O$_{32}$ $\equiv$ [OIII]$\lambda$5007/[OII]$\lambda$3727 $\gtrsim$ 5 \citep{NakajimaOuchi2014,Izotov2018a,Faisst2016,deBarros2016}. Hence O$_{32}$ is broadly used as an important diagnostic for identifying ionizing photon leakers. However, there are several other galaxies that exhibit high O$_{32}$ value, showing little or no escape of ionizing photons \citep[e.g.,][]{Stasinska2015,Bassett2019,Nakajima2020,Izotov2020}. A similar fact has also been noticed by \citet{Katz2020} in a population of simulated high-redshift galaxies. Such an apparent inconsistency could be explained by introducing the effects of neutral H{\sc i} gas porosity in the galaxy's interstellar medium (ISM) \citep{Gazagnes2020} and viewing angle \citep{Cen2015,Nakajima2020}. These effects allow the ionizing photons to escape the galaxy only from a certain region, along a particular line-of-sight, and thereby constraining the escape fraction.

It is thought that an extremely high value of O$_{32}$ in star-forming galaxies indicates the presence of either an ionization-bounded H{\sc ii} region with intense ionizing radiation produced in a massive young star formation or a completely ionized density-bounded H{\sc ii} region or a mixture of both \citep[e.g.,][]{NakajimaOuchi2014,Stasinska2015,Izotov2017}. A substantial fraction of ionizing photons is expected to escape from the density-bounded regions, as these regions are optically thin to these photons \citep{Pellegrini2012,Gazagnes2020}. The correlation between Lyman Continuum (LyC) escape fraction ($f^{esc}_{LyC}$) and $V_{sep}$ \citep{Verhamme2015,Verhamme2017}, where $V_{sep}$ denotes the peak velocity difference between the blue and the red profiles of Ly$\alpha$ emission line, has indirectly provided such evidence. According to this correlation, systems with a high $V_{sep}$ have a low $f^{esc}_{LyC}$ and vice versa. Because the high $V_{sep}$ is caused by resonant scattering of Ly$\alpha$ photons in a outflowing neutral gas, the high $f^{esc}_{LyC}$ and low $V_{sep}$ indicate a very low neutral gas column density or the presence of density-bounded regions in a leaky system. Furthermore, a similar interpretation has also been provided using nebular Mg{\sc ii}-doublet emission line, analogous to the resonant Ly$\alpha$ \citep{Henry2018}. Recently, a weak positive correlation has been found between [SII]-deficiency and $f^{esc}_{LyC}$ \citep{Wang2021}; where a high [SII]-deficiency can be interpreted as a consequence of the presence of a large optically thin region to ionizing photons \citep{Pellegrini2012}.

It is important to note that the correlations discussed above are derived from spatially integrated observations. These observations cannot exactly discriminate between the spatial extent of a ionization and density-bounded H{\sc ii} region, or distinguish whether a galaxy hosts a mixture of both. Therefore, the spatially resolved spectroscopy for the identification of a ionization or density-bounded H{\sc ii} region is a key to better characterize the true nature of leaky, compact star-forming galaxies and route to resolve why some galaxies show high high O$_{32}$ value and low or no escape of ionizing photons \citep{Bassett2019,Nakajima2020,Izotov2020}. \citet{Pellegrini2011} have demonstrated such an approach by showing the spatially resolved morphologies of optical [SII]$\lambda$6717,6731/H$\alpha$ $\equiv$ [SII]/H$\alpha$ and [SII]$\lambda$6717,6731/[OIII]$\lambda$5007 $\equiv$ [SII]/[OIII] emission line ratios for a large number of H{\sc ii} regions in the Large Magellanic Cloud (LMC) and Small Magellanic Cloud (SMC). In their study, they have shown that [SII]-enhanced ionization fronts in ionization-bounded regions surround both the ionizing massive star cluster as well as the ionized gas. In comparison, the density-bounded regions are characterized by highly ionized gas without surrounding [SII]-enhanced ionization front, indicating the sightlines for the escape of ionizing photons. In a Blister-type morphology (i.e., presence of both the ionization and density-bounded H{\sc ii} regions), [SII]-enhanced ionization front partially covers both the ionizing massive star clusters and ionized gas, allowing the ionizing radiation to escape only in certain direction. However, such a morphological study in high-redshift compact leaky galaxies is challenging due to various technical/instrumental limitations. 

Alternatively, one could study their local counterparts known as Green Pea (GP; at $0.11 \leqslant z \leqslant 0.36$) and Blueberry (at $0.02 \leqslant z \leqslant 0.05$) galaxies using spatially resolved spectroscopy. Using data in the Sloan Digital Sky Survey \citep[SDSS;][]{York2000}, GPs and Blueberries were first discovered by \citet{Cardamone2009} and \citet{Yang2017} respectively. GPs are defined by the color space by $g - r$ $\leqslant$ 2.5 and $r - i$ $\leqslant$ -0.2, while for Blueberries, they are $g - r$ $\leqslant$ 2.5 and $r - i$ $\leqslant$ -0.2. Both systems are compact ($\textless$ $1 -2$ kpc) and characterized by high EWs of [OIII]$\lambda$5007 emission line ($500 - 2500$ \AA), high ionization parameter (O$_{32}$ $\sim$ $6 - 60$), low dust extinction (E($B - V$) $\sim$ $0 - 0.25$) and low gas-phase metallicity (12 + log(O/H) $\sim$ $7.6 - 8.25$). Although, they share common properties, Blueberries are found to have low stellar-mass (M $\sim~10^{6.5} - 10^{7.5}$~M$_{\odot}$) and low star formation rate (SFR; $0.02 - 0.5$ M$_{\odot}$~yr$^{-1}$) compared to the GPs. Nevertheless, both of these systems show a similar specific$-$SFR (sSFR) lying in the range of $10^{-7} - 10^{-9}$ yr$^{-1}$. In the literature, several of these GPs and Bluebrries have now been confirmed as Lyman Continuum (LyC) and Ly$\alpha$ leakers showing an escape fraction of ionizing photons in the range of $2 - 72$\% \citep{Borthakur2014,Jaskot2019,Leitherer2016,Izotov2016a,Izotov2016b,Izotov2018a,Izotov2018b}, like the high-redshift leaky star-forming galaxies \citep[e.g.,][]{Shapley2016,Bian2017,Vanzella2018,Fletcher2019,Mestric2020,Marques2021}.

Despite the fact that spatially resolved Integral Field Unit (IFU) observations are critical for understanding the mechanisms involved in the leakage of LyC or Ly$\alpha$ photons from GPs and Blueberries, there are a very few studies in the literature. According to several of these studies, kinematical feedback and galactic outflows caused by recent starburst events are found as one of the important factors for the leakage of LyC photons in GPs \citep[e.g.,][]{Herenz2017,Micheva2019,Bosch2019}. Other study has shown that H{\sc ii} regions with low optical depth are favorable sites for ionizing photons to escape from GPs and Blueberries \citep[e.g.,][]{Chisholm2020,Bruna2021}. These limited investigations of GPs and Blueberries using spatially-resolved IFU observations clearly show the necessity for their extension to a large number of confirmed LyC and Ly$\alpha$ leakers.  

In this paper, we investigate the spatially resolved distribution of ionization and density-bounded H{\sc ii} regions in a confirmed nearby (z $\sim$ 0.0472) Ly$\alpha$ leaker (viz. SHOC 579) using the IFU Mapping Nearby Galaxies at Apache Point Observatory \citep[MaNGA;][]{Bundy2015} survey. Previously, a detailed study, based on MaNGA IFU observation of SHOC579 by Paswan et al. (2021; under review), reveals that it is a blueberry galaxy and associated with a fain low brightness disk. SHOC 579 was also observed by the Cosmic Origins Spectrograph (COS) on $Hubble~Space~Telescope~(HST)$ and found to be emitting, $f^{esc}_{Ly\alpha} \sim 10$\% \citep{Jaskot2019} and it is the only confirmed Ly$\alpha$ leaker observed in the MaNGA survey. Our current work, focuses on understanding the physical processes that allow the escape of ionizing photons from GP and Blueberry-like sources. Throughout this work, we assume a set of flat cosmological parameters such as \textit{H$_{o}$} = 70 km s$^{-1}$ Mpc$^{-1}$, $\Omega_{m}$ = 0.3 and $\Omega_{\Lambda}$ = 0.7.

\section{Data}
\label{sec:data}

The data used in the present work mainly come from the MaNGA survey. This survey is an ongoing optical IFU observing program under the fourth generation of SDSS (SDSS-IV) \citep{Bundy2015} survey that uses BOSS spectrograph \citep{Smee2013} mounted on 2.5-m Sloan Foundation Telescope \citep{Gunn2006} at Apache Point Observatory. The survey targets to observe a total of 10,000 galaxies at redshifts 0.01 $\textless$ $z$ $\textless$ 0.15 having stellar-mass of $\textgreater$ 10$^{9}$ M$_{\sun}$ \citep{Wake2017}. The current study is performed based on the galaxy sample released in the MaNGA DR16. It contains a sample of total $\sim$ 4600 galaxies out of which one galaxy (SHOC 579) is found as confirmed Ly$\alpha$ leaking Blueberry galaxy, and hence presented here. The details of its selection are presented in Paswan et al. (2021; under review).

The observed raw datacube are first reduced and calibrated using the Data Reduction Pipeline \citep[DRP;][]{Law2016}. In this process, the calibrated spectra have a wavelength coverage of $3600 - 10300$ \AA~ with a spectral resolution of $R$ $\sim$ 2000. These calibrated datacube are then made as science-ready products using Data Analysis Pipeline \citep[DAP;][]{Westfall2019}. In this study, we use the DAP output products. The DAP uses pPXF code \citep{Cappellari2004} with MILES stellar library, having the best set of observed stellar spectra available to date, and models both the stellar continuum and line (emission and absorption) features identified in the spectra from each spaxel of datacube. In this fitting, all the line fluxes are derived with Gaussian model fit after stellar continuum subtraction, and then provided in the form of 2D maps. The above mentioned pipelines (i.e., DRP\footnote{For more detail about the DRP, see the link $-$ \url{https://svn.sdss.org/public/repo/manga/mangadrp/tags/v1_5_4/}} and DAP\footnote{For more detail about the DAP, see the link $-$ \url{https://sdss-mangadap.readthedocs.io/en/latest/}}) are SDSS software, and they are mainly designed for reducing and analysing the IFUs data observed in the MaNGA survey. 

All the emission line fluxes and their ratios used in the present work are corrected for both the Galactic and internal reddening using their corresponding $E(B - V)$ values. The Galactic reddening is first applied assuming reddening law provided by \citet{O'Donnell1994}. The internal reddening correction to the galaxy is applied using line flux ratio of f$_{H\alpha}$/f$_{H\beta}$ assuming the expected theoretical value as 2.86 and the Case-B recombination \citep{Osterbrock1989} with an electron temperature of $\sim$ 10$^{4}$ K and electron density of 100 cm$^{-3}$. For some spaxels, the line flux ratios of f$_{H\alpha}$/f$_{H\beta}$ are found to be below the expected theoretical value of 2.86. A low value of f$_{H\alpha}$/f$_{H\beta}$ is often associated with intrinsically low reddening, and hence we assumed an internal $E(B - V)$ values as zero for such cases \citep[e.g.,][]{Paswan2018,Paswan2019}. Throughout this work, we have used only those emission lines whose signal-to-noise ratio (SNR) is $\geqslant$ 3. 

\section{Results and analysis}
\label{sec:result}

\begin{figure}
\begin{center}
\rotatebox{0}{\includegraphics[width=0.6\textwidth]{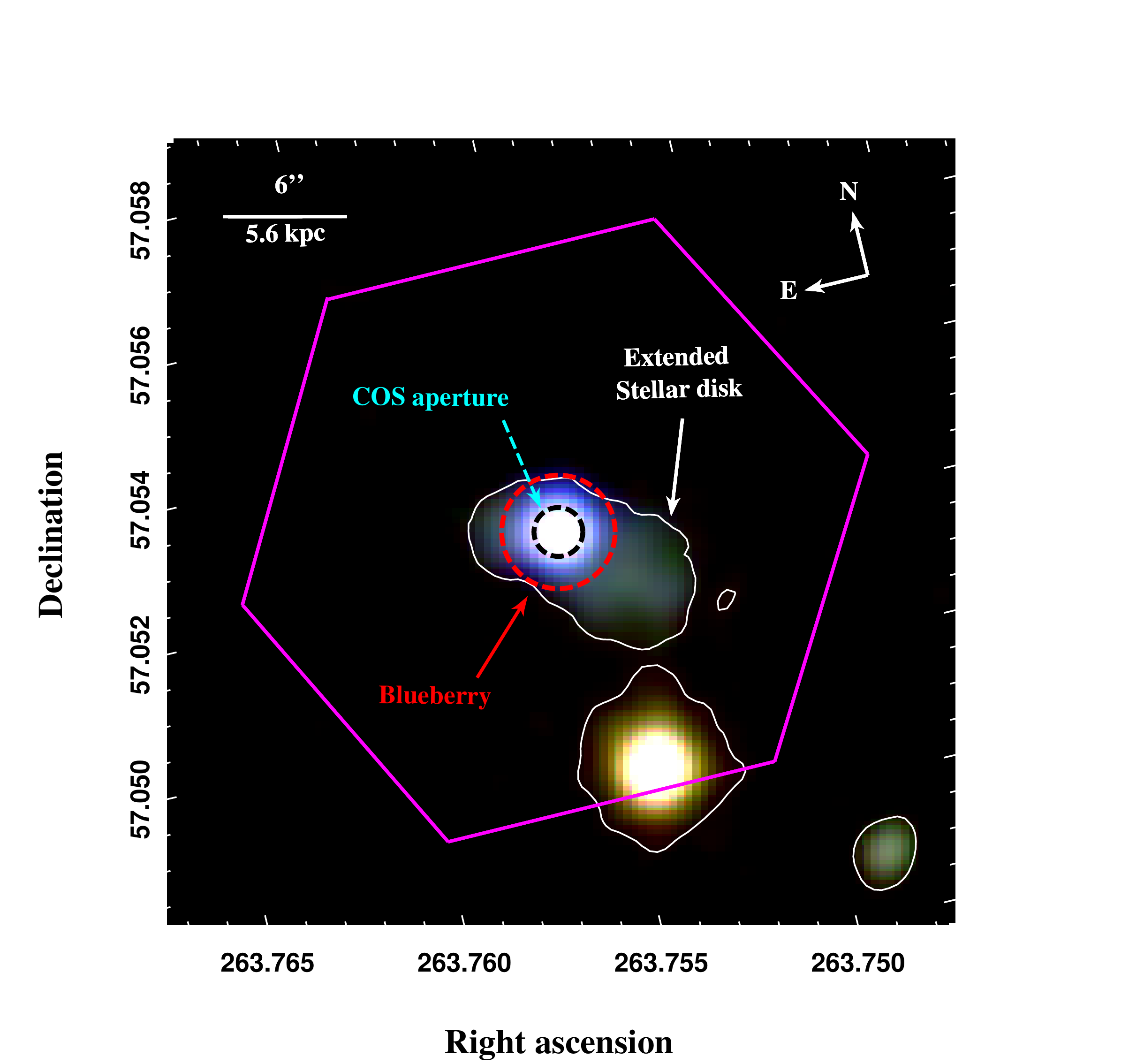}}
\caption{The SDSS $gri$-band color composite image of SHOC 579 overlaid with MaNGA footprint (magenta hexagon). Compact blue starburst region (Blueberry) and extended stellar disk of the galaxy are represented with regions belong to in and outside the red dashed circle, respectively. The white contour is drawn at 3$\sigma$ level showing the extent of extended stellar disk. The dashed black circle shows the aperture (2.5" in diameter) used in the COS observations \citep{Jaskot2019}.}
\label{color-comp}
\end{center}
\end{figure}

The SDSS $gri$-band color composite image of SHOC 579 overlaid with the MaNGA footprint is shown in Fig.~\ref{color-comp}. The basic properties of SHOC 579 obtained from the literature are listed in Table~\ref{Table1}. 

\begin{table}
\centering
\caption{Properties of the studied galaxy.}
\vspace {0.3cm}
\begin{tabular}{cc} \hline 
Parameter                  &	Value  \\\hline
MaNGA ID                   &  $8626 - 12704$ \\
RA                         &  17$^{h}$ 35$^{m}$ 01.25$^{s}$ \\ 
DEC                        &  +57$^{d}$ 03$^{m}$ 09$^{s}$\\
Redshift (z)               &   0.0472 \\
Angular Size               &  14.4 $\times$ 10.2 (arcsec$^{2}$) \\
Linear size                &  13.3 (kpc) \\
Stellar-mass               &  3.15 $\times$ 10$^{9}$ (M$_{\odot}$)\tablenotemark{\small a} \\
SFR                        &  9.74 $\pm$ 0.15 (M$_{\odot}$~yr$^{-1}$)\tablenotemark{\small b} \\
12 + log(O/H)              &  8.11 $\pm$ 0.07\tablenotemark{\small b} \\
EW(H$\alpha$)              &  1442 $\pm$ 9 (\AA)\tablenotemark{\small b}\\
O$_{32}$                   &  6.8 $\pm$ 0.3\tablenotemark{\small b} \\
$E(B - V)$\tablenotemark{\small c}                 &  0.12 $\pm$ 0.01\tablenotemark{\small b} \\
EW(Ly$\alpha$)             &  64 $\pm$ 4 (\AA)\tablenotemark{\small c}\\
$V_{sep}$            &  460 $\pm$ 47 (km~s$^{-1}$)\tablenotemark{\small c}\\
$f^{esc}_{Ly\alpha}$ & 0.09 $\pm$ 0.01\tablenotemark{\small c} \\
\hline 
\label{Table1}
\end{tabular}
\flushleft
\tablenotetext{\small a}{Taken from Paswan et al. (2021; under review).}
\tablenotetext{\small b}{Values taken from \citet{Jaskot2019} which are derived using SDSS fiber slit spectroscopy.}
\tablenotetext{\small c}{Internal reddening, after corrected for Milky Way extinction.}

\end{table}

\subsection{Probing density-bounded regions}
\label{DB-IB}

\begin{figure*}
\begin{center}
\rotatebox{0}{\includegraphics[width=0.49\textwidth]{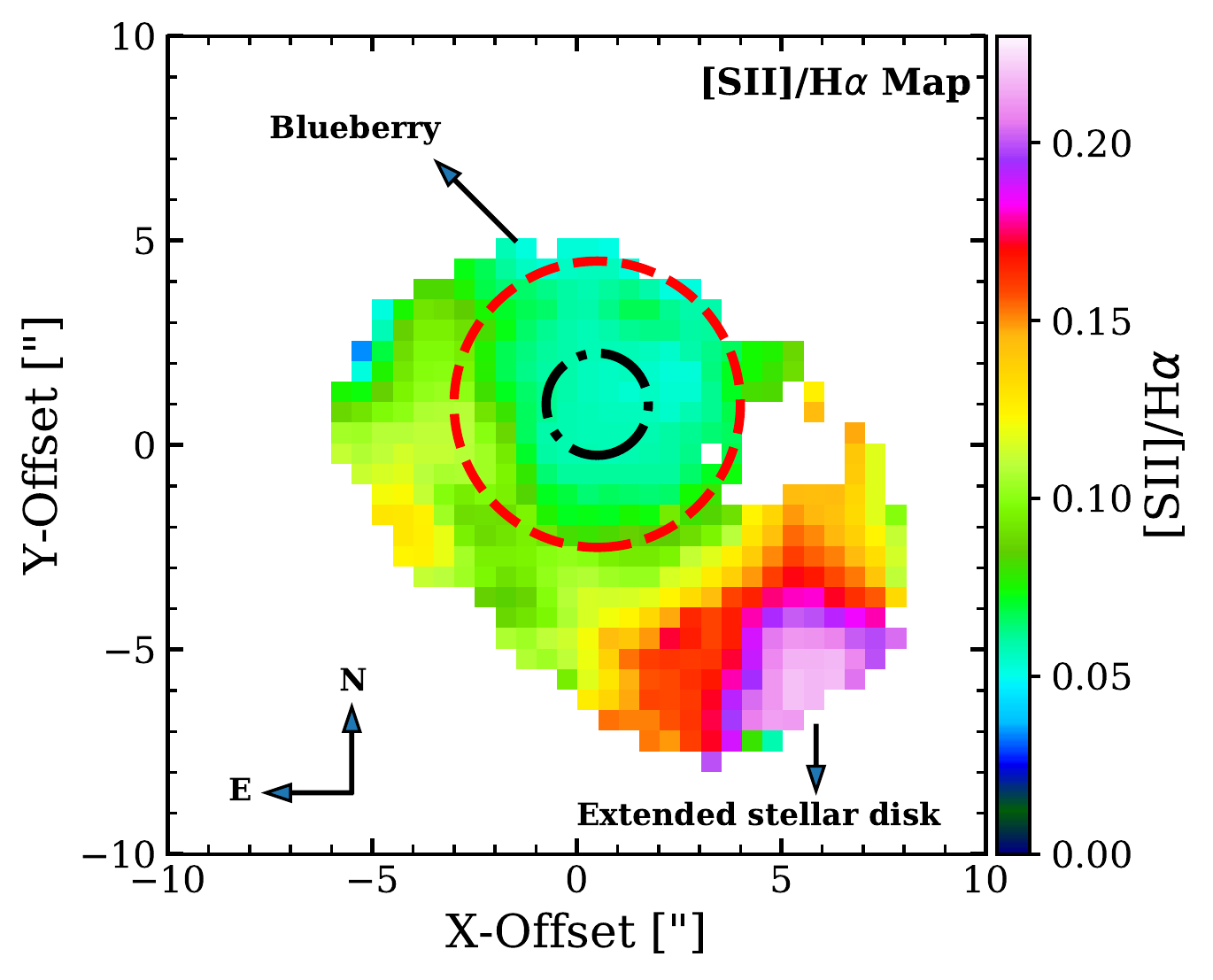}}
\rotatebox{0}{\includegraphics[width=0.49\textwidth]{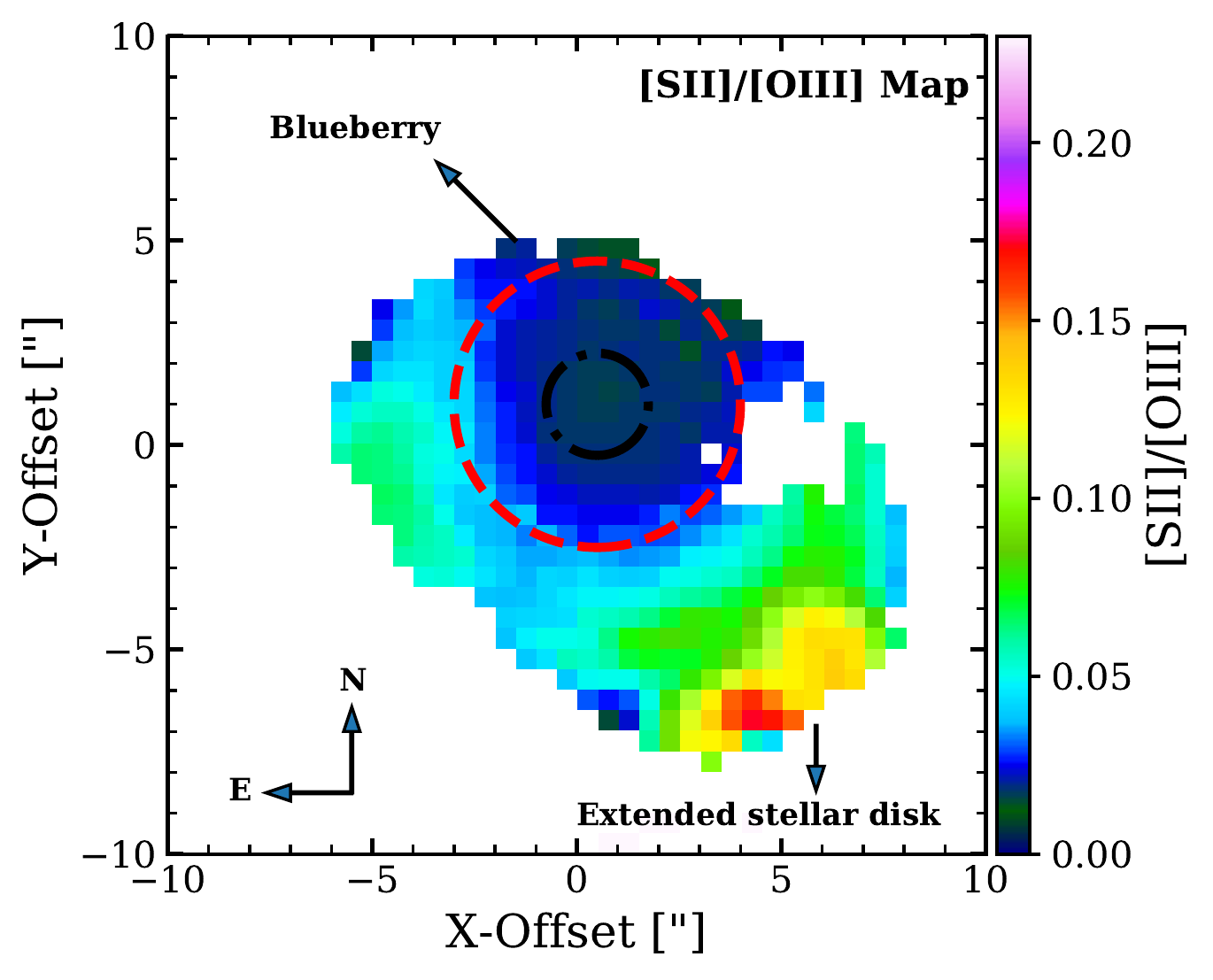}}
\caption{The spatially resolved 2D-map of [SII]/H$\alpha$ (left) and [SII]/[OIII] (right) emission line ratios. The drawn red and black circles in both the panels denote the same Blueberry region and COS-aperture as explained in Fig.~\ref{color-comp}.}
\label{em-morph}
\end{center}
\end{figure*}

One of the main results of this work is highlighted in Fig.~\ref{em-morph}. It represents the spatially resolved morphologies of optical [SII]/H$\alpha$ (left panel) and [SII]/[OIII] (right panel) emission line ratios used for distinguishing the density and ionization-bounded H{\sc ii} regions. It works on the following principle: the enhancement or depletion in the intensity of [SII]$\lambda$6717,6731 relative to H$\alpha$ and [OIII]$\lambda$5007 emission lines is a consequence of the interplay between different ionization potentials of S$^{o}$, H$^{o}$ and O$^{+}$ elements present in the galaxy's ISM and ionizing photons emitted from young ($\leqslant$ 5 Myr) massive ($\geqslant$ 10 M$\odot$) stars. The ionization potential for producing [SII]$\lambda$6717,6731 (i.e., 10.4 eV) is significantly less than a Rydberg (13.6 eV), implying that much of [SII]$\lambda$6717,6731 emission emerges in the warm and partially ionized region just near and beyond the outer edge of the H{\sc ii} regions. In those H{\sc ii} regions which are density-bounded (i.e., optically thin to ionizing photons), such partially ionized [SII] zones appear very weak or even absent \citep[e.g.,][]{Pellegrini2011,Pellegrini2012}. As a result, the intensity of [SII]$\lambda$6717,6731 relative to H$\alpha$ and [OIII]$\lambda$5007 emission lines declines significantly.  

Following the above discussion, we noticed that the intensity of [SII]$\lambda$6717,6731 relative to H$\alpha$ and [OIII]$\lambda$5007 emission lines drops to a very low value within the central starburst region (i.e., inside the red circle in Fig.~\ref{em-morph}), implying this region as density-bounded or optically thin to ionizing photons. In contrast, there is a significant [SII]-enhancement towards South-West of the galaxy and a partial [SII]-enhancement in a arc-like morphology towards the North-East outskirt, implying these regions as as optically thick to ionizing photons (i.e., ionization-bounded regions). The median values of [SII]/H$\alpha$ and [SII]/[OIII] ratios in the blueberry region are 0.06 and 0.02, respectively, while they are found as 0.12 and 0.06 (i.e., a factor of $2 - 3$ enhanced) in the extended disk region. Overall, the morphology of [SII]/H$\alpha$ and [SII]/[OIII] emission line ratios suggests that SHOC 579 hosts a Blister-type H{\sc ii} regions, opened for leakage of ionizing photons only through certain density-bounded/optically-thin region seen towards North and North-West regions.

\subsection{Spatially resolved [SII]-deficiency}
\label{sec:SII}

In this section, we analyse the spatially resolved distribution of [SII]-deficient spaxels in SHOC 579. We follow the definition of [SII]-deficiency as in \citet{Wang2019}. In that, [SII]-deficiency is the difference between the observed value of log([SII]/H$\alpha$) and its inferred value obtained using the empirical relation given by:

\begin{eqnarray}
    y = 0.487 + 0.014\eta + 0.028\eta^{2} - 0.785\eta^{3} - 3.870\eta^{4}  \\ \nonumber + 0.446\eta^{5} + 8.696\eta^{6} + 0.302\eta^{7} - 6.623\eta^{8},
    \label{eq:SIIdef}
\end{eqnarray}

\begin{figure*}
\begin{center}
\rotatebox{0}{\includegraphics[width=0.51\textwidth]{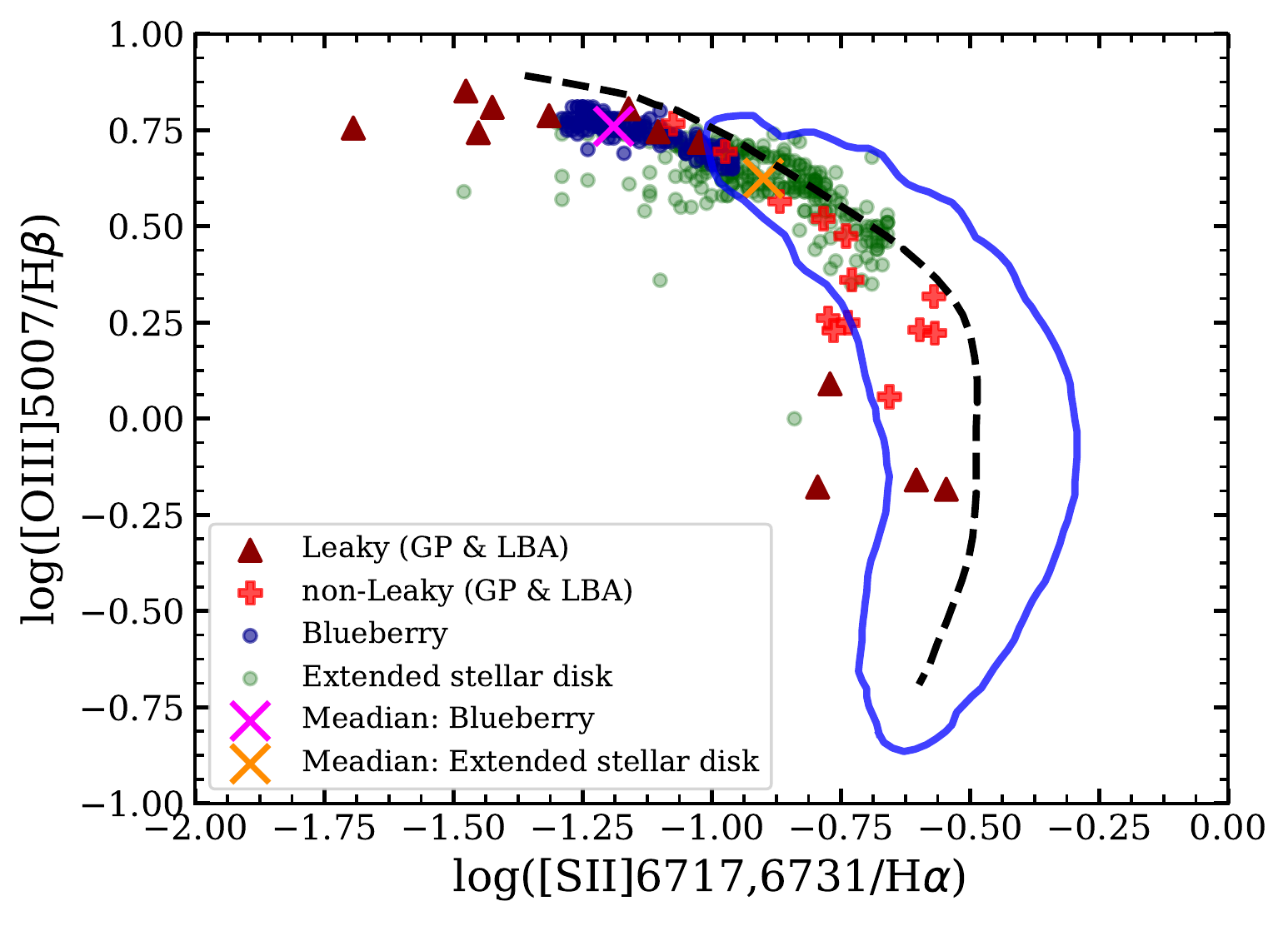}}
\rotatebox{0}{\includegraphics[width=0.48\textwidth]{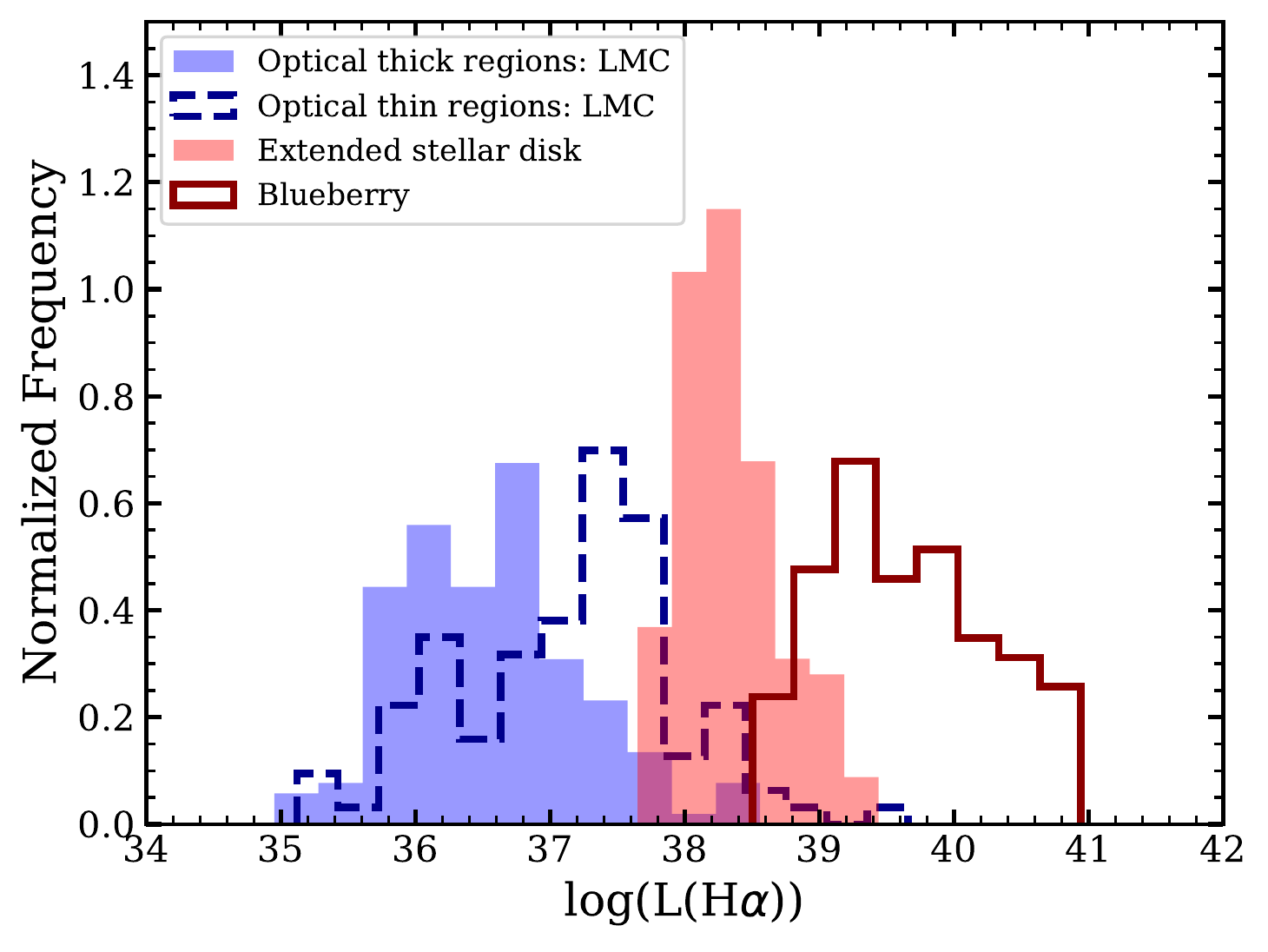}}
\caption{Left: In order to define [SII]-deficiency, the [SII]$\lambda$6717,6731/H$\alpha$ emission line ratio is plotted against [OIII]5007/H$\beta$ emission line ratio. In this plot, solid contour (light-blue) shows the density distribution enclosing 90\% of typical star-forming galaxy sample observed in the SDSS DR12. The black dashed line denotes the locus of peak density of star-forming galaxies. Here, the [SII]-deficiency is defined as the horizontally galaxy displacement in log([SII]$\lambda$6717,6731/H$\alpha$) from the reference black dashed line. Triangle and plus symbols represent leaky and non-leaky objects, respectively, taken from a sample of GPs \citep{Izotov2016a,Izotov2016b,Izotov2018a,Izotov2018b} and LBAs \citep{Alexandroff2015}. Blue and green dots represent the spaxels belonging to the central starburst region (inside the circle) and extended stellar disk regions (outside the circle), respectively, as shown in Fig.~\ref{em-morph}. Right: the H$\alpha$ luminosity distribution of optically thick (blue filled) and thin (blue dashed line) H{\sc ii} regions in LMC. Similarly, the the H$\alpha$ luminosity distribution of [SII]-deficit or optically thin (solid red line) and [SII]-enhanced or optically thick (red filled) spaxels in the galaxy.}
\label{SII-def-LHa}
\end{center}
\end{figure*}

In the above equation, y $\equiv$ log([SII]/H$\alpha$) is the reference value for a given observed value of $\eta$ $\equiv$ log([OIII]/H$\beta$). This empirical relation, denoted as a dashed line in Fig.~\ref{SII-def-LHa}, is obtained after fitting a polynomial to the peak density of typical star-forming galaxies falling on the plane defined by log([OIII]/H$\beta$) and log([SII]/H$\alpha$) line ratios.

In the left panel of Fig.~\ref{SII-def-LHa}, it can be seen that most of the leaky galaxies are found to be relatively more [SII]-deficient (i.e., located at larger horizontal distance from the reference black dashed line) compared to non-leaky systems which fall inside the solid contour (i.e., closer to reference black dashed line) defined for typical star-forming galaxies. Similarly, we also find spaxels from the density-bounded region in SHOC 579 (blue dots) are consistent with the location of previously known leaky systems, showing a relatively large [SII]-deficiency. In contrast, most of the spaxels belonging to the ionization-bounded regions (green dots; except a few spaxels in the vicinity of blueberry region) fall inside the contour or closer to the reference black dashed line, suggesting that they are relatively less [SII]-deficient compared to spaxels in the blueberry region. Ly$\alpha$ photons, unlike LyC photons, can escape through the optically thick or ionization-bounded region because they are involved in a resonant scattering process. However, the density-bounded region allows Ly$\alpha$ photons to escape rather easily \citep[][and references therein]{Huan2017}. We observe that the COS aperture used in \citet{Jaskot2019} for UV observations only covers the density-bounded region of SHOC 579 (see Fig.~\ref{color-comp}). This suggests that the observed escape of Ly$\alpha$ photons by \citet{Jaskot2019} was from a straightforward route, with little to no impediment to the leakage process.

The right panel in Fig.~\ref{SII-def-LHa} shows the distribution of H$\alpha$ luminosity of optically thin (blue dashed line) and thick (blue filled) H{\sc ii} regions in LMC \citep{Pellegrini2012}. This distribution is bimodel in nature, indicating that the H$\alpha$ luminosity distribution for optically thick H{\sc ii} regions peaks at lower value than those of optically thin H{\sc ii} regions. A similar bimodal distribution of H$\alpha$ luminosity is also observed in SHOC 579, where the luminosity for density-bounded (i.e., optically thin) H{\sc ii} regions (solid red line) peaks at higher luminosity compared to the optically thick H{\sc ii} regions (red filled). These results suggest that the H$\alpha$ luminosity of density-bounded H{\sc ii} region in general dominates over the luminosity of ionization-bounded H{\sc ii} region. This is most likely due to the presence of a relatively stronger starburst event in the density-bounded region.

\subsection{Reconciling the observed $f^{esc}_{Ly\alpha}$ with the spatial dust extinction}
\label{recon}

\begin{figure*}
\begin{center}
\rotatebox{0}{\includegraphics[width=0.49\textwidth]{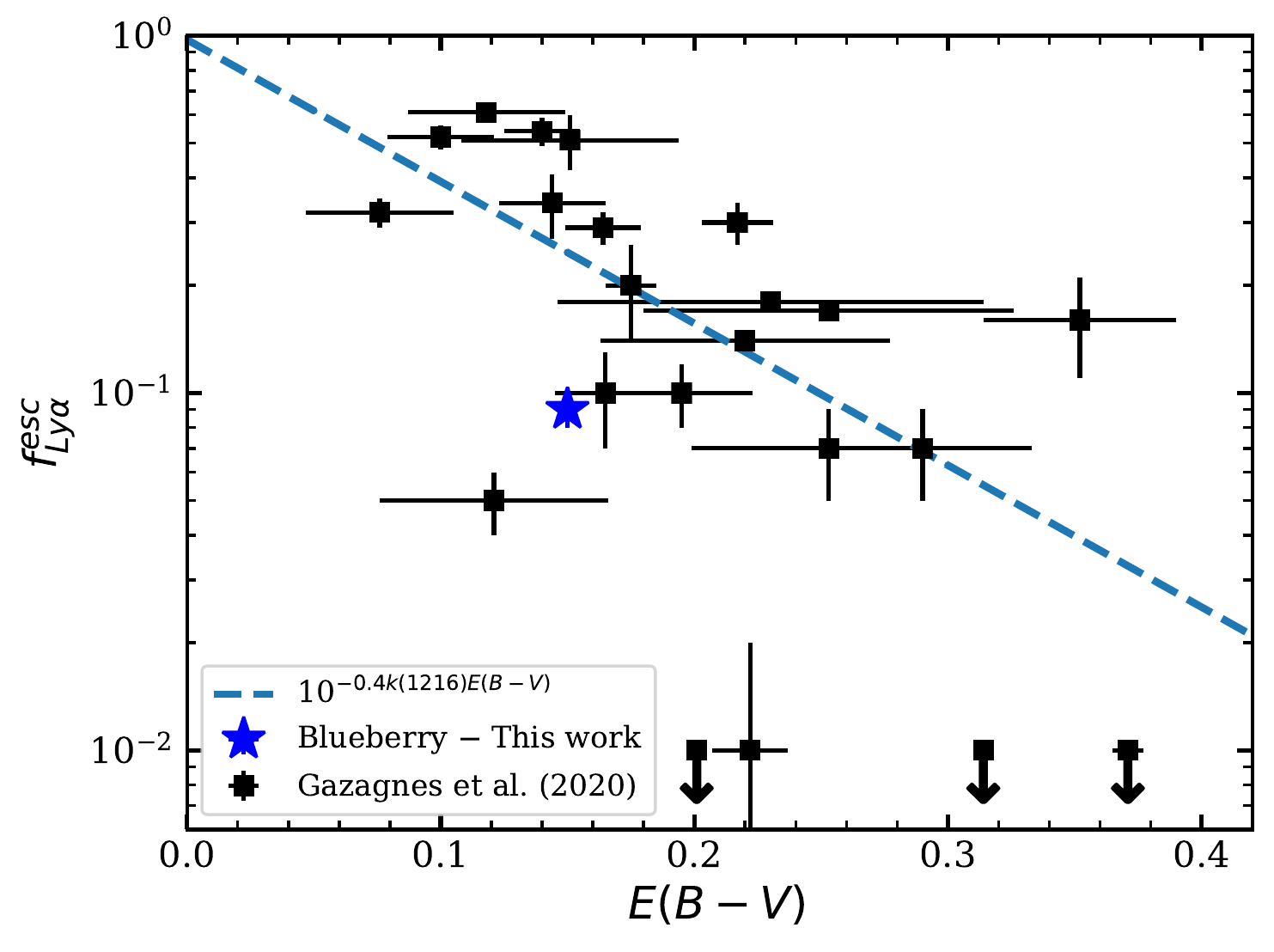}}
\caption{Relation between the observed $f^{esc}_{Ly\alpha}$ and $E(B - V)$ with logarithmic y-scale. Solid square data are taken from \citet{Gazagnes2020}, and dashed line represents the shape of the dust attenuation curve at the wavelength of Ly$\alpha$ emission line i.e., 1216 \AA. Star and solid circle represent the data in the present work. Here star and solid circle show the dust extinction averaged over Blueberry (i.e., inside the dashed circle) and extended stellar disk (i.e., outside the dashed circle), respectively. The value of $f^{esc}_{Ly\alpha}$ is used as 0.09 $\pm$ 0.01 \citep[see][]{Jaskot2019}.}
\label{EBV-flya}
\end{center}
\end{figure*}

\begin{figure*}
\begin{center}
\rotatebox{0}{\includegraphics[width=0.49\textwidth]{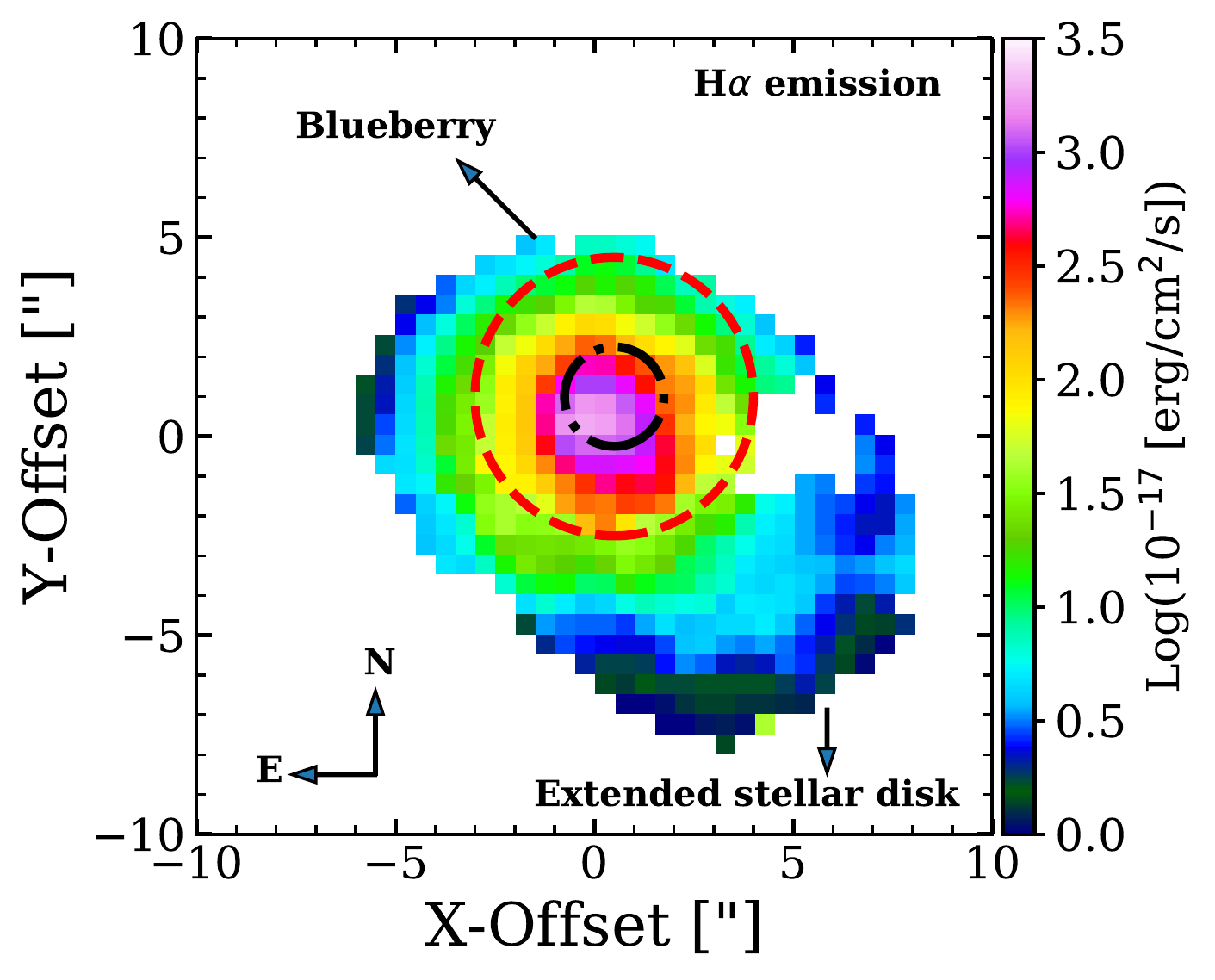}}
\rotatebox{0}{\includegraphics[width=0.49\textwidth]{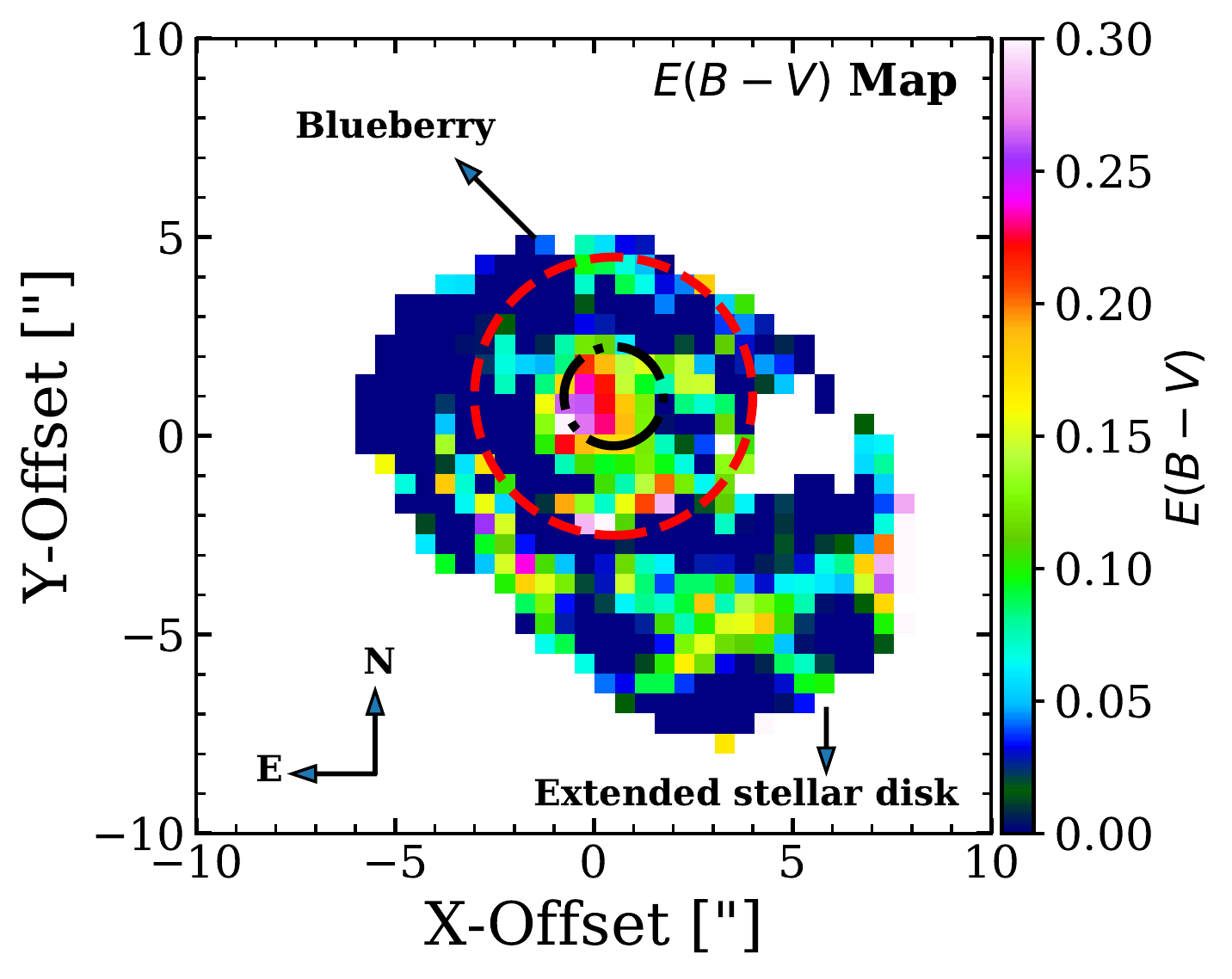}}
\caption{The 2D spatially resolved H$\alpha$ emission line map (left) corrected for the dust reddening, E($B - V$), as presented by 2D-map in the right panel. This reddening map is derived by following the Balmer decrement method. The drawn red and black circles in both the panels denote the same Blueberry region and COS-aperture as explained in Fig.~\ref{color-comp}.}
\label{halpha-ext}
\end{center}
\end{figure*}

Since SHOC 579 is dominated by a strong starburst (SFR $\sim$ 17 M$_{\odot}$~yr$^{-1}$) with a moderate ionization parameter (O$_{32}$ $\gtrsim$ 7; see Sect.~\ref{siii-sii}), the physical conditions are therefore likely to favour a huge production of Ly$\alpha$ photons that could easily escape through the density-bounded region. However, the attenuation of Ly$\alpha$ photons due to dust extinction plays a crucial role in determining its escape fraction. As can be seen in Fig.~\ref{EBV-flya}, $f^{esc}_{Ly\alpha}$ decreases as nebular dust extinction $E(B - V)$ increases. Moreover, similar evidence can be found in other studies published in the literature \citep[e.g.,][]{Atek2009,Hayes2015,Huan2017}. We measure an average $E(B-V) \simeq 0.15$ for SHOC 579 within the COS aperture as shown in Fig.~\ref{em-morph}. Based on this value of dust extinction and the fitted attenuation relation by \citet{Gazagnes2020}, as shown in Fig.~\ref{EBV-flya}, we expect $f^{esc}_{Ly\alpha} \geq$ 20\% for our galaxy. This expected value of $f^{esc}_{Ly\alpha}$ is also consistent with the Ly$\alpha$ escape fraction obtained using empirical relation that uses measured $V_{sep}$ \citep[][see Eq. 5]{Izotov2020MNRAS}, where $V_{sep}$ for the galaxy in the present study is used as $\sim$ 460~km~s$^{-1}$ (see Table~\ref{Table1}). However, the directly measured $f^{esc}_{Ly\alpha} \sim$10\% \citep{Jaskot2019} falls below these expected values (see Fig.~\ref{EBV-flya}). We explain this observed disparity using our spatially resolved H$\alpha$ emission line map (corrected for dust extinction) and its related dust extinction map as shown in the left and right panels of Fig.~\ref{halpha-ext}, respectively. The density-bounded region covered under the COS-aperture (i.e., the [SII]-deficit region within the black circle depicted in Fig.~\ref{em-morph}) exhibits a strong H$\alpha$ line emission, suggesting the predicted location of massive Ly$\alpha$ photon generation. This region, however, is dominated in part by a considerable amount of dust with a colour excess of $E(B - V)$ $\sim 0.1 - 0.2$ ($\sim$ 0.15 on an average). On the spatially resolved scale, it can be seen that the West side of the COS aperture seems to have $E(B - V) \sim 0.1$, whereas the East side have $\textgreater 0.2$. A large number of Ly$\alpha$ photons can be destroyed by such a dust distribution, and only a small proportion of generated Ly$\alpha$ photons can escape via the porous dusty region (i.e., probably from the West side of the COS aperture). Thus, previously reported $f^{esc}_{Ly\alpha}$ $\sim$10\% by \citet{Jaskot2019} is likely to be limited by the porous dust distribution in SHOC 579. This implies that the dust porosity can constrain the escape fraction of Ly$\alpha$ photons, similar to how the neutral gas porosity constrains the escape fraction of LyC photons \citep{Gazagnes2020}. 

\subsection{Correlation between LyC escape fraction and [SII]-deficiency} \label{sec:correlation}

\begin{figure*}
\rotatebox{0}{\includegraphics[width=0.49\textwidth]{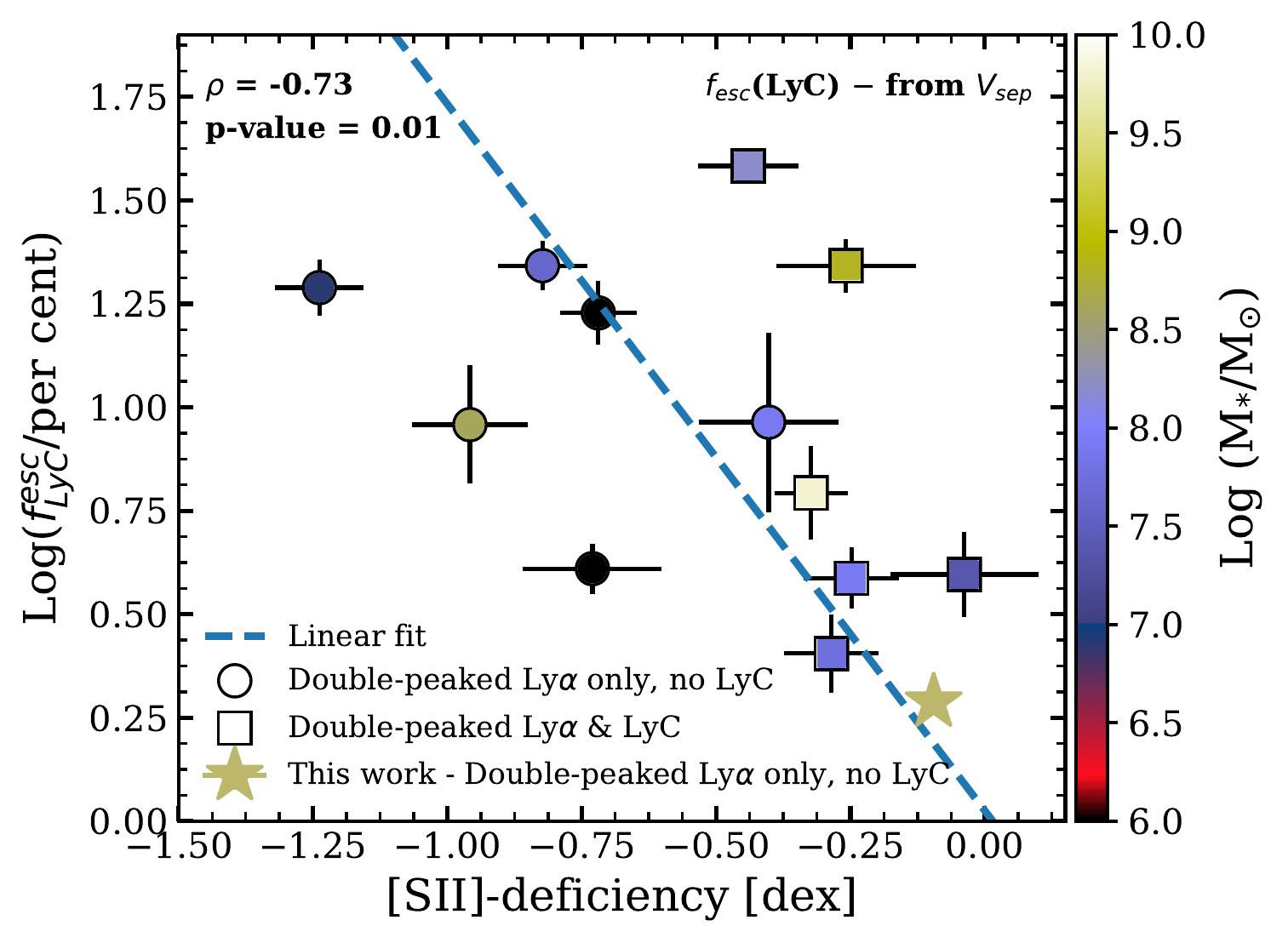}}
\rotatebox{0}{\includegraphics[width=0.48\textwidth]{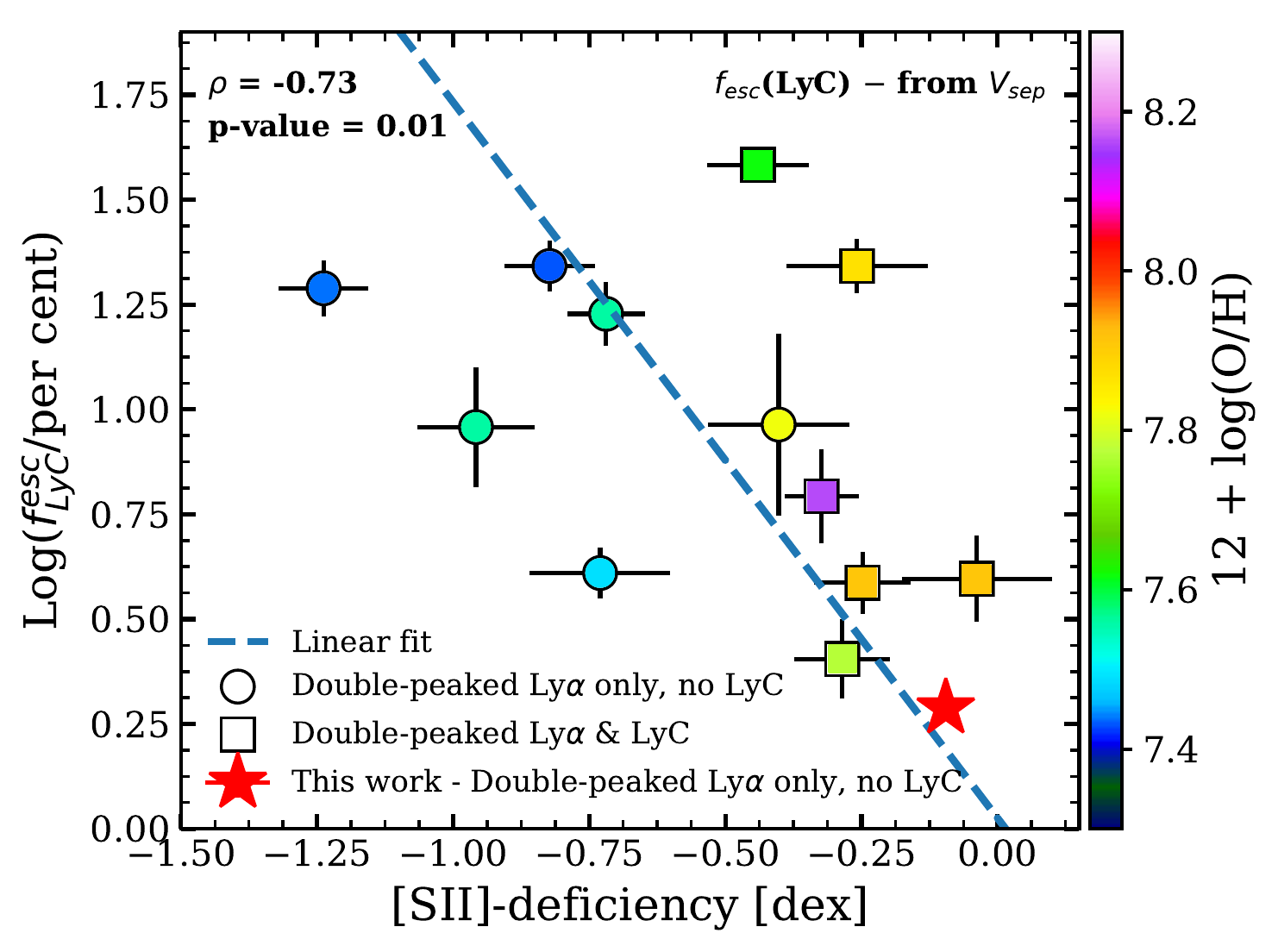}}
\rotatebox{0}{\includegraphics[width=0.5\textwidth]{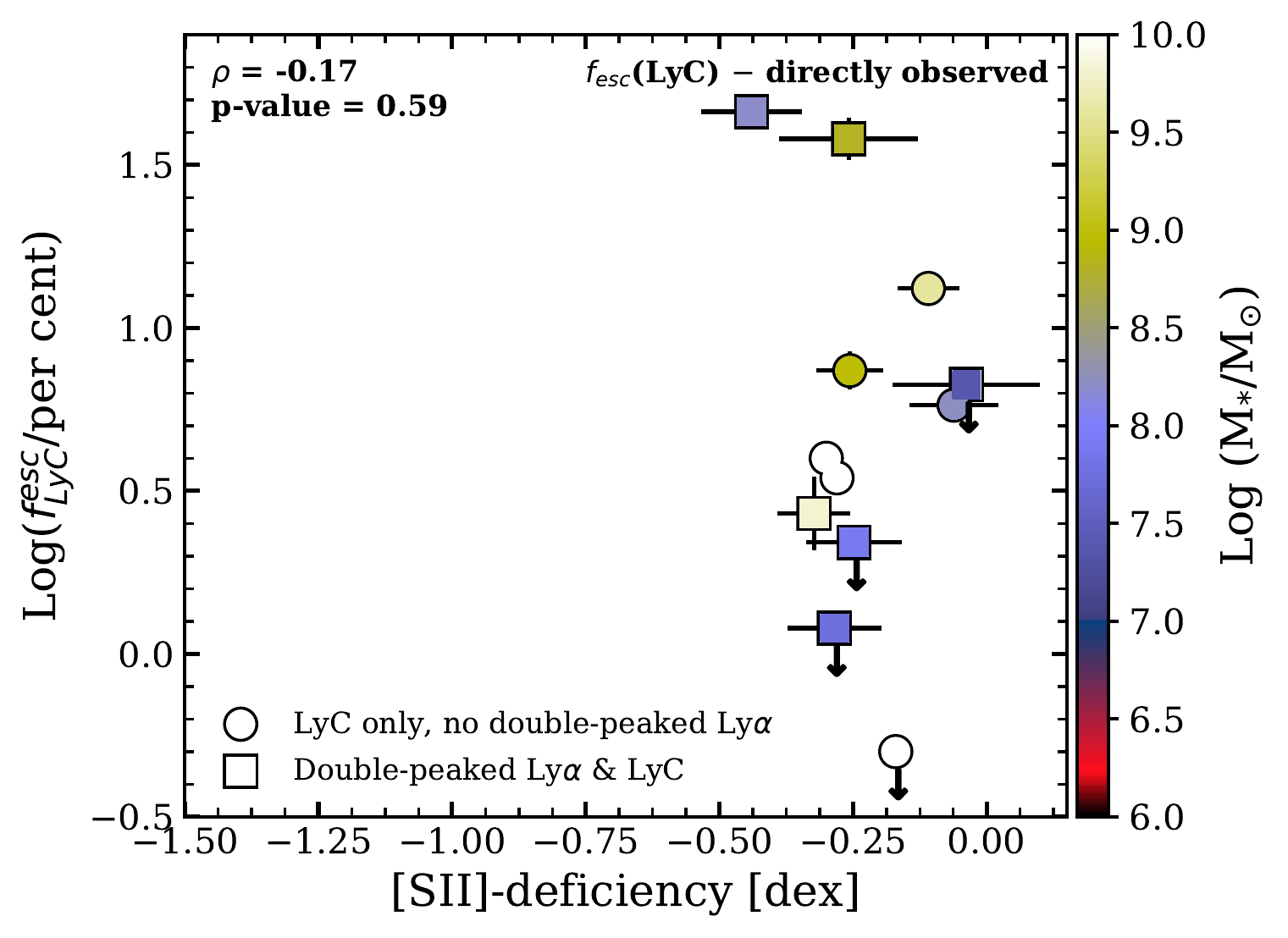}}
\rotatebox{0}{\includegraphics[width=0.49\textwidth]{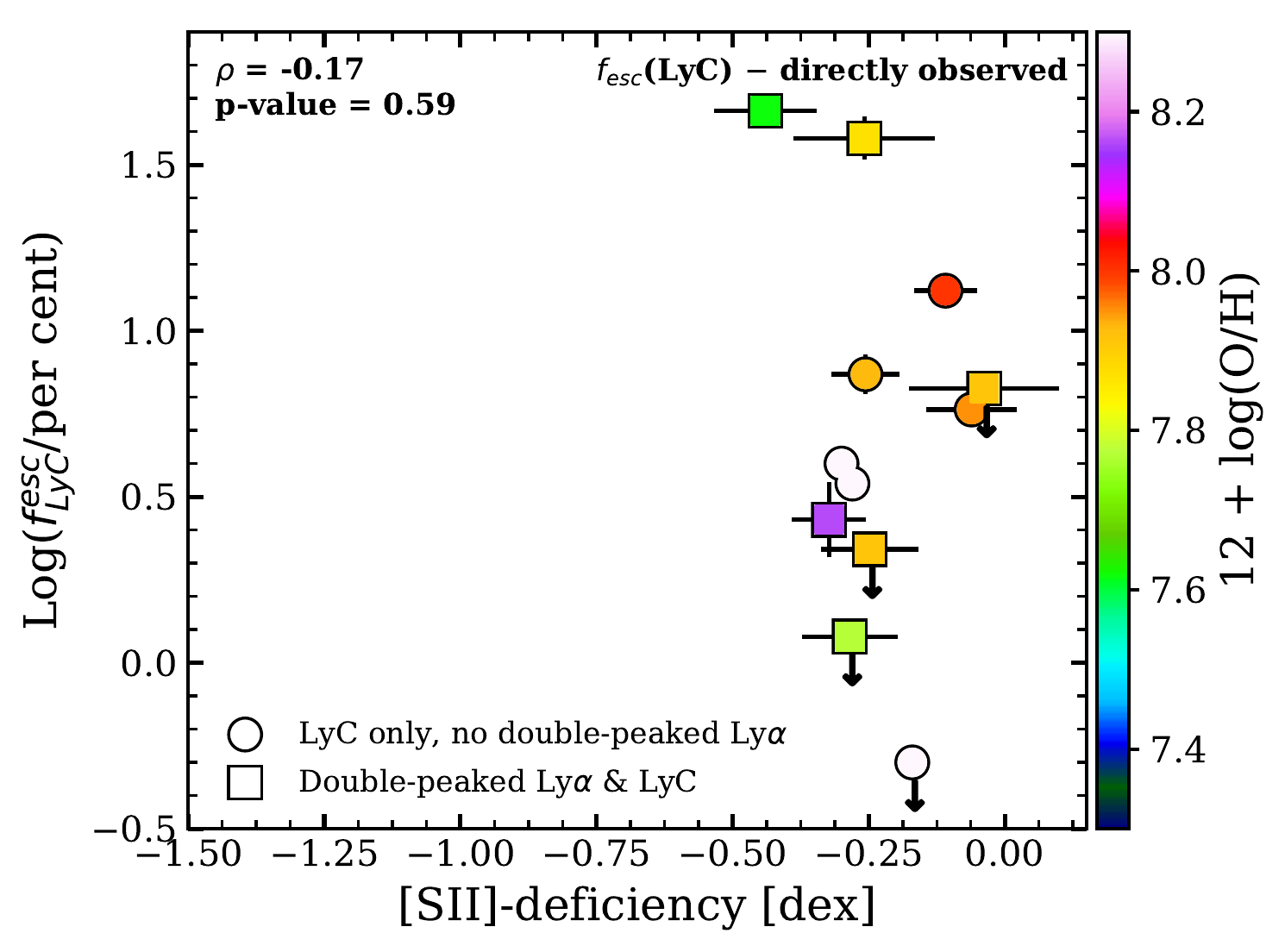}}
\caption{The plots displaying correlation between LyC escape fraction and [SII]-deficiency for a sample of GPs/Blueberries taken from \citet{Izotov2016a,Izotov2016b,Izotov2018a,Izotov2018b,Izotov2020,Izotov2021} and \citet{Wang2019}. The top two panels are shown for galaxies whose $f^{esc}_{LyC}$ is inferred though $V_{sep}$, while the same in the bottom two panels is for galaxies having direct measurements of $f^{esc}_{LyC}$. The circles represent GPs/Blueberries having only double-peaked Ly$\alpha$ emission line and no LyC photons as shown in upper panels, and showing only LyC photons with no double-peaked (single peak) Ly$\alpha$ emission line as shown in lower panels. In both the upper and lower panels, the squares represents those GPs/Blueberries which have been observed with having double-peaked Ly$\alpha$ emission line along with LyC photons. The star denotes the data for the galaxy in the present work. The colorbars in the left and right columns represent the stellar mass and gas-phase metallicity of galaxies, respectively. The dashed straight lines in the top two panels are error-weighted linear fits to the galaxies.}  
\label{SII-def-LyC-esp}
\end{figure*}

Based on evidence found in the literature \citep{Pellegrini2011,Pellegrini2012,Wang2019}, including the current work, it is now clear that [SII]-deficient H{\sc ii} regions are favourable sites for ionizing photon escape. Therefore, using a sample of 18 confirmed Ly$\alpha$ and LyC leaky GPs/Blueberries as shown in Fig~\ref{SII-def-LyC-esp}, we explore the relation between $f^{esc}_{LyC}$ and [SII]-deficiency. The confirmed sample of leaky sources is drawn from the literature \citep[e.g.,][]{Izotov2016a,Izotov2016b,Izotov2018a,Izotov2018b,Wang2019,Izotov2020,Izotov2021}. Some galaxies in this sample, including SHOC 579, do not have a direct measurement of $f^{esc}_{LyC}$ but the Ly$\alpha$ lines with $V_{sep}$ are observed. The empirical relation between $V_{sep}$ and $f^{esc}_{LyC}$ reported by \citet{Izotov2018b} is used to infer the values of $f^{esc}_{LyC}$ for such sources. The inferred $f^{esc}_{LyC}$ for SHOC 579 is 1.94 $\pm$ 0.05 \%. Out of 18 sources, a total of 6 sources have both the directly measured and inferred values of $f^{esc}_{LyC}$. The top two panels in Fig.~\ref{SII-def-LyC-esp} are shown for galaxies whose $f^{esc}_{LyC}$ is inferred using $V_{sep}$, whereas the bottom two panels are shown for galaxies with direct $f^{esc}_{LyC}$ measurements. 

A moderately significant correlation between inferred $f^{esc}_{LyC}$ and [SII]-deficiency can be found in the top panels of Fig.~\ref{SII-def-LyC-esp}, suggesting that galaxies with higher [SII]-deficiency have larger $f^{esc}_{LyC}$. Interestingly, the inferred $f^{esc}_{LyC}$ $\sim$ 2\% for a measured [SII]-deficiency in SHOC 579 is in accordance with this observed correlation. On the other hand, we see significant scatter when directly measured $f^{esc}_{LyC}$ is made to correlate with [SII]-deficiency (see bottom panels of Fig.~\ref{SII-def-LyC-esp}). This is likely due to a very narrow range of [SII]-deficiency measured for leaky galaxies in our sample. It is possible that we might see an improved correlation after increasing the sample size having a broad dynamic range of [SII]-deficiency. In fact, \citet{Wang2021} have recently demonstrated a slightly improved correlation between directly observed $f^{esc}_{LyC}$ and [SII]-deficiency using a sample of 35 low-$z$ LyC leakers. In their galaxy sample, although the values of [SII]-deficiency vary between 0.1 and -0.8, with 6 sources lying between -0.6 and -0.8; their derived correlation is still weak. In order to have a better understanding, a large sample of confirmed LyC leakers, in particular, at higher end of [SII]-deficiency is perhaps required. Furthermore, we notice that a substantial scatter in the correlation is seen at lower end of [SII]-deficiency (i.e., between -0.4 and 0) in our sample, consistent with the scatter observed by \citet[][see their Fig. 4]{Wang2021}. Following their argument presented in \citet[][see their Fig. 5]{Wang2021}, it appears that less [SII]-deficient galaxies would have a large covering fraction of neutral H{\sc i} gas surrounding the ionized gas. In such a ISM structure, LyC photons are expected to escape via low-column-density channels or hole(s) \citep[e.g.,][]{Borthakur2014,Rivera-Thorsen2015,Gazagnes2018}. In addition, the escape of LyC photons depends on observer's line-of-sight. Such a scenario suggests that an amount of observed escape fraction of LyC photons in galaxies with low [SII]-deficiency would be highly uncertain. Therefore, the observed substantial scatter in the correlation, at lower end of [SII]-deficiency, might be attributable to line-of-sight variation caused by porous H{\sc ii} regions. In the literature, a few observational studies of galaxies showing high O$_{32}$ but low $f^{esc}_{LyC}$ have already been found to be consistent with the hypothesis of line-of-sight variation \citep[e.g.,][]{Steidel2018,Nakajima2020}. Additionally, the observed scatter might also be attributable to anisotropically escaping LyC photons \citep[e.g.,][] {Zastrow2011,Cen2015}.

In all panels of Fig.~\ref{SII-def-LyC-esp}, we bring the stellar mass and gas-phase metallicity of the leaky systems while discussing these correlations. We find that $f^{esc}_{LyC}$ and [SII]-deficiency correlation is unlikely to depend on the host stellar mass, consistent with earlier findings \citep{Izotov2021,Saxena2021}. Moreover, this also supports the finding of \citet{Wang2021} that there is no link between [SII]-deficiency and stellar-mass.  However, the observed correlation (upper panels of Fig.~\ref{SII-def-LyC-esp}) appears to be dependent on the gas-phase metallicity of leaky systems. Sources with lower gas-phase metallicity have higher $f^{esc}_{LyC}$ and higher [SII]-deficiency. This inference indeed supports the fact that compact metal-poor starburst galaxies are one of the favourable sources contributing to the reionization of the early Universe \citep[e.g.][]{Ouchi2009,Mitra2013,Naidu2018,Kimm2019}.
\\ 

\section{Wolf-Rayet feature and physical conditions of ionized gas}\label{WRf}

\begin{figure*}
\begin{center}
\rotatebox{0}{\includegraphics[width=0.68\textwidth]{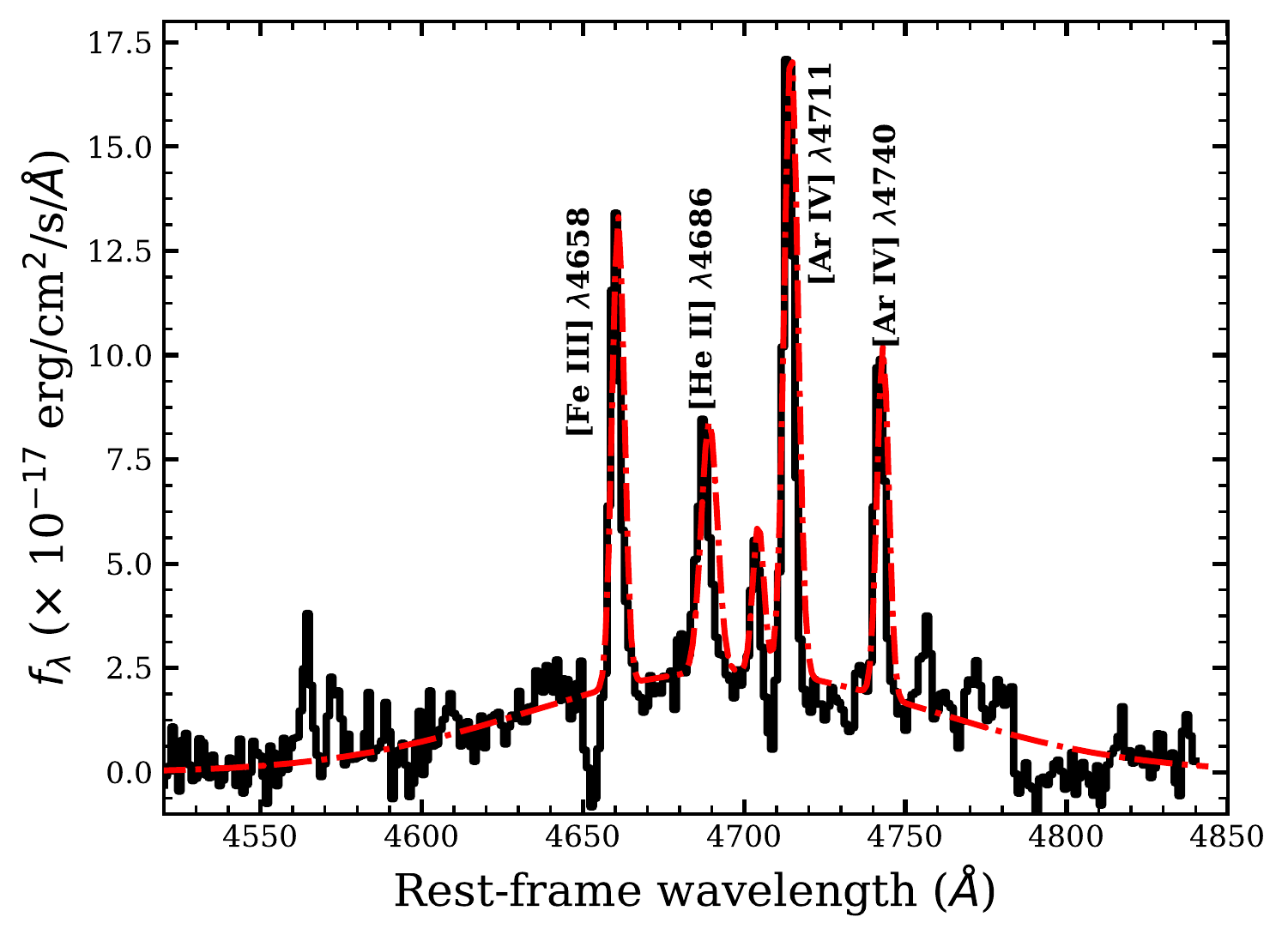}}
\caption{The detection of WR bump feature along with He {\sc ii} $\lambda$4686 and other WR emission lines. All the emission lines are marked and labeled. The red dashed line represents the Gaussian models fitted to the narrow emission lines, including broad WR feature.}  
\label{WR-feat}
\end{center}
\end{figure*}

In Fig.~\ref{WR-feat}, we show the detection of a broad Wolf-Rayet (WR) bump feature along with nebular He {\sc ii}$\lambda$4686,  [Fe {\sc iii}] $\lambda$4658, and [Ar {\sc iv}] $\lambda$4711,4740 emission lines, derived after stacking all the spectra from extreme starburst region (i.e., inside the red circle) of the BBLAE in SHOC 579. The presence of the WR feature indicates a substantial population of WR stars and a fairly recent ($\la 5-6$ Myr) intense starburst event in this region. No WR feature is detected in the extended stellar disk region. The nebular He {\sc ii}$\lambda$4686 line, which is sometimes seen in conjunction with WR features, shows the presence of relatively hard ionizing radiation, probably produced by young and hot WR stars or other mechanisms \citep[cf.][]{Shirazi2012}. So far, there seems to be no indication for the causal relation between the presence/absence of nebular He {\sc ii} and escape of LyC photons \citep[e.g.][]{Guseva2020}.

\begin{figure*}
\begin{center}
\rotatebox{0}{\includegraphics[width=0.48\textwidth]{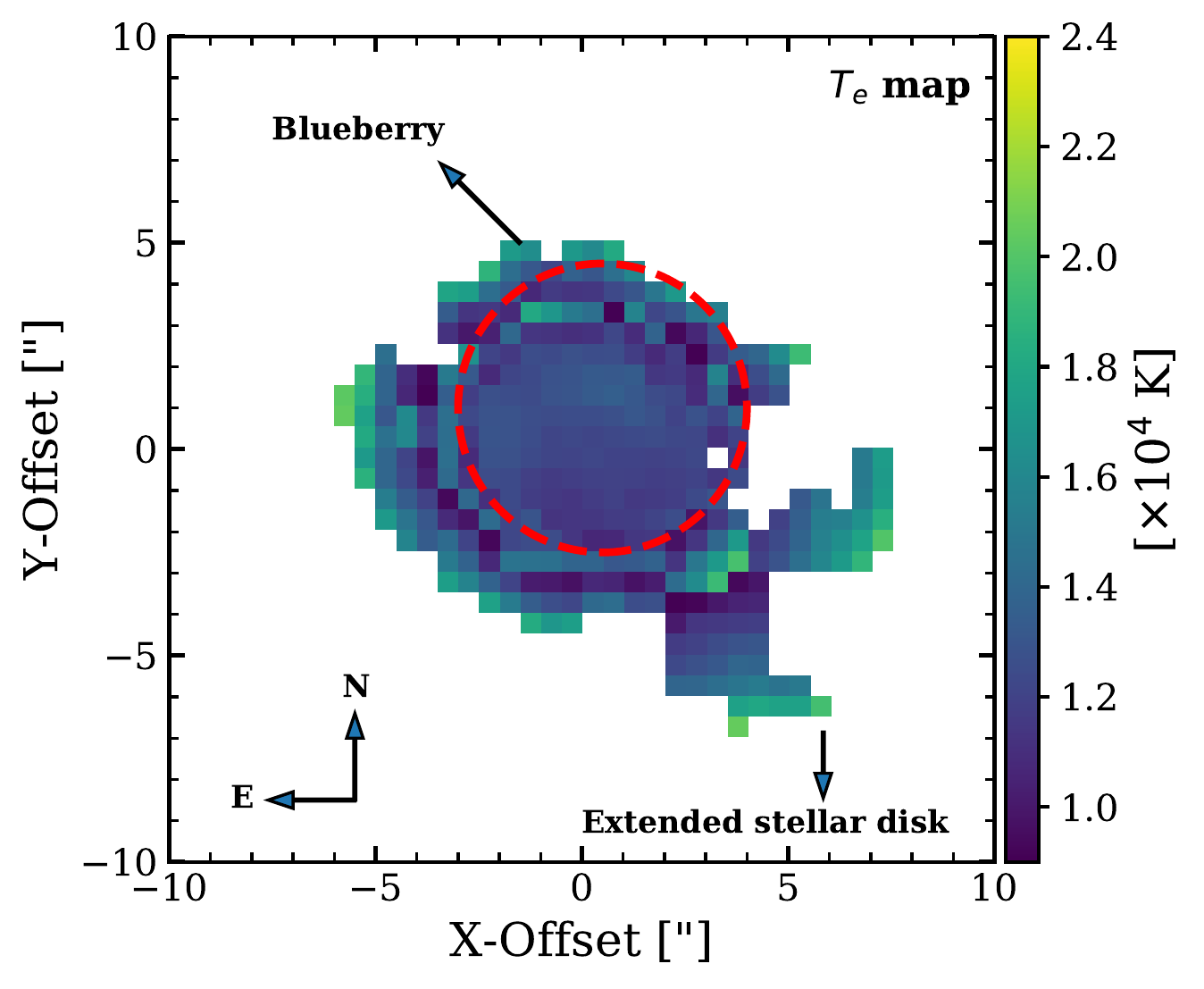}}
\rotatebox{0}{\includegraphics[width=0.48\textwidth]{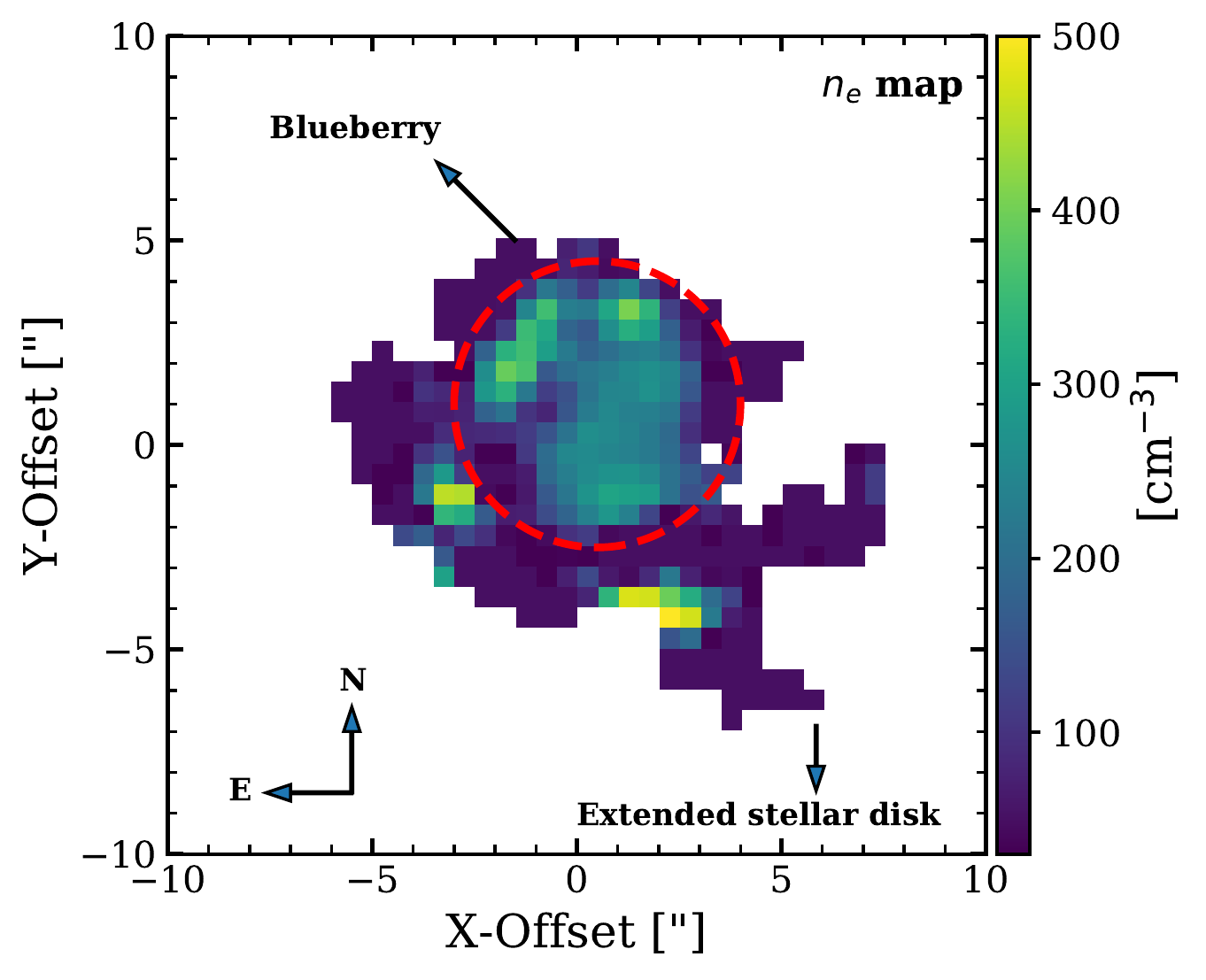}}
\caption{The spatially resolved 2D-map of electron temperature $T_{e}$ (K) and density $n_{e}$ (cm$^{-3}$) in the galaxy. The drawn circle in each panel has the same meaning as explained for Blueberry region in Fig.~\ref{color-comp}.}
\label{e-den}
\end{center}
\end{figure*}

Fig.~\ref{e-den} depicts the spatially resolved maps of electron temperature ($T_{e}$; left panel) and density ($n_{e}$; right panel) of the ionized gas in SHOC 579. Here $T_{e}$ was estimated using the temperature diagnostic ratio of [OIII]$\lambda$4959+[OIII]$\lambda$5007/[OIII]$\lambda$4363 by utilising the five-level program in the {\sc iraf nebular} task \citep{Shaw1995}, because the weak auroral [OIII]$\lambda$4363 emission line is observed from many spaxels of the galaxy. The electron density in each spaxel (for their estimated $T_{e}$ in this work) is then calculated using the same five-level algorithm and the nebular [SII]$\lambda$6717,6731 doublet emission line. As seen in Fig.~\ref{e-den}, in comparison to the rest of the galaxy (i.e., out side the circle), the density-bounded region within the circle has a considerably higher electron density. The higher electron density, particularly in the density-bounded H{\sc ii} region, is consistent with typical electron density obtained in prior low-redshift LyC/Ly$\alpha$ leakers such as GPs \citep[e.g.,][]{Izotov2020,Guseva2020}. In the ionization-bounded H{\sc ii} region (i.e., extended disk outside the circle), it is typical ($\leqslant$ 100 cm$^{-3}$), comparable to that observed in normal star-forming galaxies \citep[e.g.,][]{Guseva2011,Izotov2011,Paswan2018,Paswan2019}.

\section{[S{\sc iii}]/[S{\sc ii}]: a proxy of ionization parameter}
\label{siii-sii}

\begin{figure*}
\begin{center}
\rotatebox{0}{\includegraphics[width=0.465\textwidth]{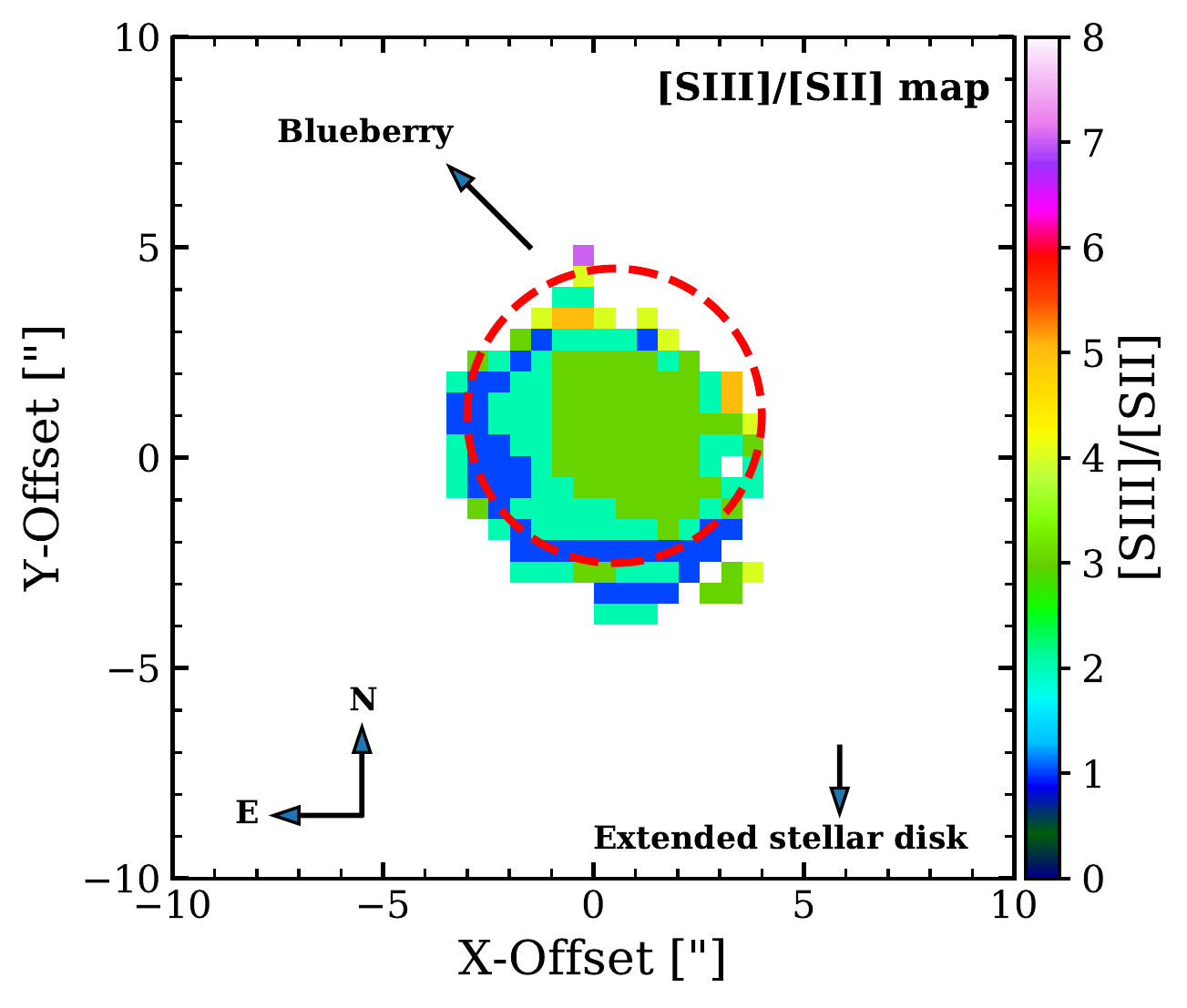}}
\rotatebox{0}{\includegraphics[width=0.52\textwidth]{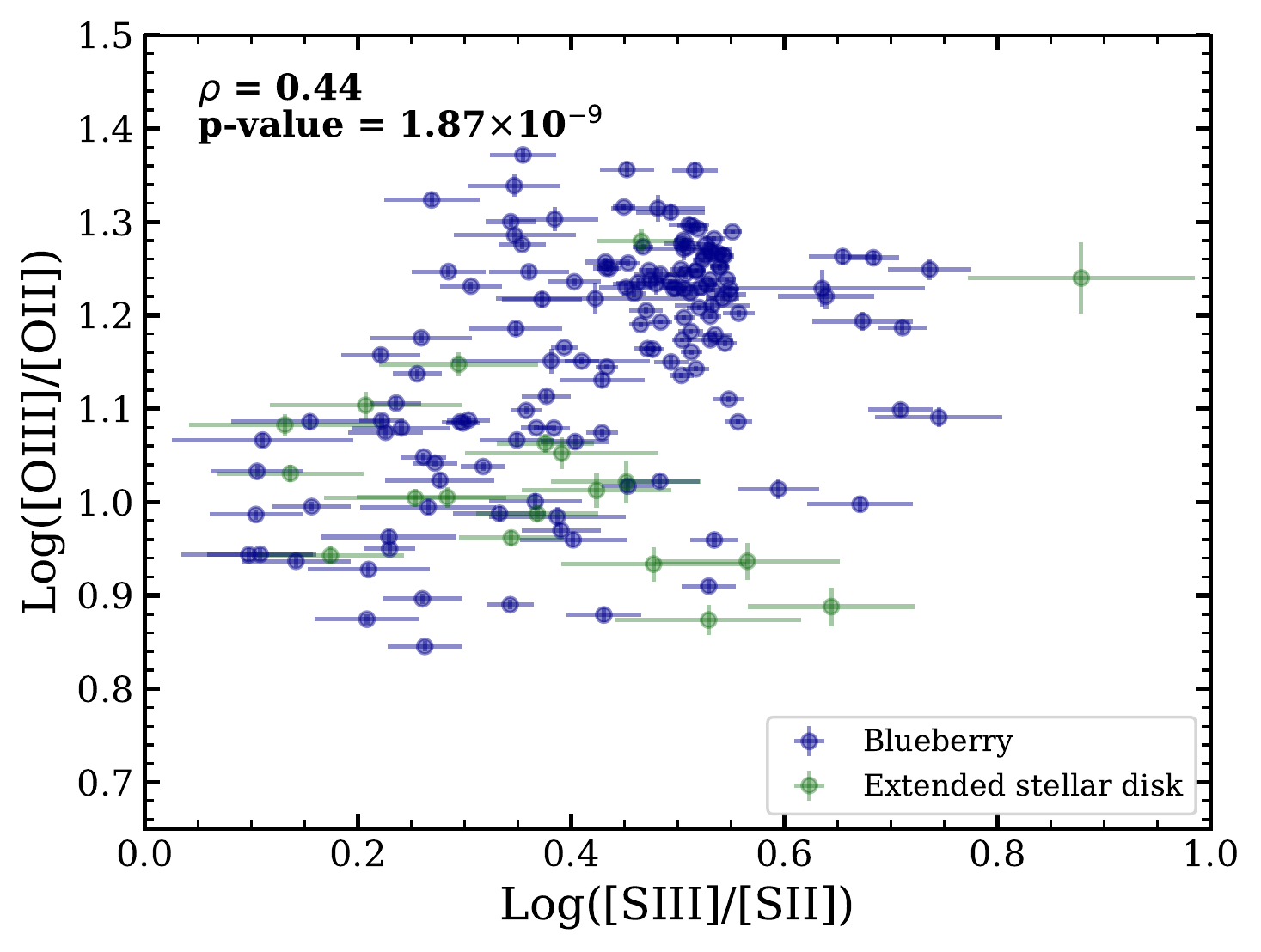}}
\caption{The spatially resolved 2D-map of S$_{32}$ parameter (left panel) and its spatially resolved correlation with O$_{32}$ parameter (right panel). The spatial values of O$_{32}$ parameter used here  are taken from the previous study by Paswan et al. (2021; under review). The drawn circle in the left panel has the same meaning as explained for Blueberry region in Fig.~\ref{color-comp}.} 
\label{SIII-ana}
\end{center}
\end{figure*}

[SIII]$\lambda$9069,9532/[SII]$\lambda$6717,6731 $\equiv$ S$_{32}$ is used as a proxy for the ionization parameter \citep[e.g.,][]{Diaz1985,Sanders2019}, like O$_{32}$ parameter \citep[e.g.,][]{Baldwin1981,Alloin1978}. While previous studies have used integrated spectra recorded within the slit aperture, we investigate their spatially resolved behaviour in our BBLAE as shown by the 2D map in Fig.~\ref{SIII-ana} (see left panel). The [SIII]$\lambda$9069,9532 emission line fluxes used for this map are estimated using our own fit to the emission line profiles rather than pPXF (see Appendix~\ref{SIII-emana} for more details). It was not feasible to derive S$_{32}$ map throughout the extended stellar disc region due to low SNR ($\textless$ 3) of [SIII]$\lambda$9069,9532 emission lines (except for a few spaxels on the margins of the circle depicted in the left panel of Fig.~\ref{SIII-ana}). However, [SIII]$\lambda$9069,9532 emission line can be clearly seen in the density-bounded/starburst region (within dashed circle). We note that the values of S$_{32}$ parameter are higher in the inner region of the starburst event than the outer region, consistent with O$_{32}$ map as shown by Paswan et al. (2021; under review). 

The spatially resolved connection between S$_{32}$ and O$_{32}$ parameters is shown in Fig.~\ref{SIII-ana} (right panel), and it indicates a weak positive correlation ($\rho = 0.44$). This finding, on the other hand, suggests a tendency of rising O$_{32}$ with increasing S$_{32}$, which is consistent with prior research \citep[e.g.,][]{Diaz1985,Sanders2019}. In Fig.~\ref{SIII-ana} (left panel), most of the spaxels in the Blueberry region of the galaxy have a value of S$_{32}$ $\textgreater$ 2, equating to an value of O$_{32}$ $\textgreater$ 10 or so. This inference further implies that starburst region in our BBLAE indeed shows a intense and hard radiation field, consistent with the high O$_{32}$ parameter and the detection of nebular He{\sc ii} $\lambda$4686 emission line. 

\section{Discussion}
\label{sec:discuss}

In the present work, we investigate the presence of density or ionization-bounded H{\sc ii} regions, or a mixture of both, in a confirmed LAE, SHOC 579, at $z \sim$ 0.047. These regions are potential locations for leakage, non-leakage, or partial leakage of ionizing photons. In our investigation, we construct the spatially resolved morphologies of optical [SII]/H$\alpha$ and [SII]/[OIII] emission line ratios using MaNGA IFU data and provide insights into the structural state of galaxy's ISM that in turn plays a critical role in interpreting the leakage of ionizing and Ly$\alpha$ photons from SHOC 579. We found that only the central starburst region in the galaxy (as shown by the red circle in Fig.~\ref{em-morph}) is [SII]-deficient (see the left panel of Fig.~\ref{SII-def-LHa}), i.e. density-bounded, and thus is likely to leak ionizing photons. The rest of the galaxy is ionization-bounded, i.e. [SII]-enhanced, and unlikely to leak ionizing photons. Our analysis indicates that the previously reported small escape fraction of Ly$\alpha$ photons of 10 \% is from density-bounded H{\sc ii} region in SHOC 579, and it is mostly constrained by a significant amount of porous dust. Furthermore, we also find a moderately significant correlation between $f^{esc}_{LyC}$ inferred through $V_{sep}$ and [SII]-deficiency.

It is well-understood that starburst regions, escaping Ly$\alpha$ photons, are often associated with dust \citep{Atek2009,Hayes2015,Axel2020}. The distribution of these dust around starburst regions may be dense and/or porous, as also seen in the present study (see Fig.~\ref{halpha-ext}). In future, it would be interesting to investigate the relationship between dust covering fraction and $f^{esc}_{Ly\alpha}$ using IFU-based spectroscopic observations of a larger sample of confirmed Ly$\alpha$ emitter, similar to the work published by \citet{Gazagnes2020} who showed the correlation between covering fraction of H{\sc i} gas and $f^{esc}_{LyC}$ for leaky system. A similar argument may also be used to limit the escape fraction of ionizing photons, because dust also decreases the $f^{esc}_{LyC}$ \citep[e.g., see Fig. 12 in][]{Gazagnes2020}. Furthermore, it is worth noting that the aperture used in the HST/COS-observation covers only a part of the [SII]-deficient or density-bounded region in SHOC 579. Aperture size limitation and positioning across the expected region of escaping Ly$\alpha$ or LyC photons may result in a lower limit to the escape fraction of these photons. Therefore, identifying the spatial location of density and ionization-bounded H{\sc ii} regions in potential LAEs and leaky systems, before planning the UV-observations, might be a crucial step forward to better constraining the Ly$\alpha$ and LyC escape fraction. 

In the literature, previously reported correlations between $f^{esc}_{LyC}$ and different observable parameters such as O$_{32}$ \citep{Huan2017,Izotov2021}, $V_{sep}$ of resonant Ly$\alpha$ \citep{Verhamme2015,Verhamme2017,Izotov2021} and nebular Mg{\sc ii}-doublet emission line \citep{Henry2018}, dust extinction \citep{Huan2017,Gazagnes2020} and [SII]-deficiency \citep{Wang2021} have been performed using spatially unresolved observations of leaky systems with slit-based aperture. Note that all these correlations show significant scatter. Perhaps, a part of the scatter is contributed due to spatially unresolved observations that show some impacts owing to global measurements of aforementioned observable parameters. The spatially resolved studies of these correlations with IFU observations might reduce the scatter. 

In the recent past, there has been a significant progress in our understanding of the leaky systems, especially when IFU observations are used. \citet{Herenz2017} found ionised gas filaments in the circumgalactic halo of SBS 0335-52E, an extremely metal-poor starburst galaxy, using VLT/MUSE IFU data. They claim that the filaments discovered are potential pathways for LyC photons to escape. Furthermore, using IFU-like fiber-fed Potsdam Multi-Aperture Spectrophotomete (PMAS) observations, \citet{Micheva2019} investigated the kinematics and physical characteristics of the ionised gas in the nearest LyC emitter galaxy, Mrk 71/NGC 2366. They demonstrated that kinematic feedback from the starburst event causes an optically thin region, implying that the escape fraction of LyC photons along the gas outflows is non-zero. \citet{Bruna2021} recently used VLT/MUSE data to depict the ISM state in NGC 7793 at $\sim$ 10 pc scale. Using the S[II]/[OIII] ratio as a proxy for the optical depth of the gas, they constrain the escape fraction of ionizing photons after categorising H{\sc ii} regions into ionization bounded or featuring channels of optically thin gas, similar to our current work. 

In light of the above, \citet{Chisholm2020} published a spatially resolved morphological analysis of the Mg{\sc ii}-doublet emission line ratio ($R$ = $F_{2796}/F_{2803}$) in a leaky GP galaxy, J1503+3644, at z = 0.36, using data from the Keck{\sc ii} telescope's image slicer Integral Field Spectrograph (IFS). The optically thin H{\sc ii} regions responsible for the reported $f^{esc}_{LyC}$ of 6\% were mapped using the $R$ values. Note that the Mg{\sc ii}-doublet emission line ratio-based morphology is only applicable for sources with redshifts greater than 0.2, as most optical ground-based spectroscopic surveys (e.g., MaNGA, MUSE/VLT, and KCWI Keck, among others) have wavelength coverage ranging from $\sim$ 3300 to 10000 \AA. However, for low-redshift sources such as Blueberries (0.02 $\leqslant$ z $\leqslant$ 0.05) and GPs at z $\leqslant$ 0.2, the [SII]-deficiency based morphological method may be a better indirect diagnostic for detecting LyC leakers. It may also be used to study high-redshift starburst galaxies using infrared (IR) spectroscopic data, in addition to low-redshift sources. However, due to the significant contribution of sky background in the IR waveband, IR spectroscopy is difficult and needs either a large aperture ground or spaced-based telescope. In this context, the $HST$ IR WFC3/G141 slitless grism observations in the 3D-$HST$ survey \citep[see,][]{Momcheva2016} might be beneficial for examining the [SII]-deficiency in high redshift compact starburst galaxies. Similarly, data from upcoming mega space and ground-based observing facilities, such as $James~Webb~Space~Telescope$ (JWST), TMT and ELT, may be essential for distant starburst galaxies in the early Universe. 

\section{Conclusions}

The main conclusions of this study are as follows.

{\begin{itemize}
    \item SHOC 579 hosts a mixture of density and ionization-bounded H{\sc ii} regions $-$ indicating a Blister-type morphology. Our analysis indicates that the Ly$\alpha$ photons \citep{Jaskot2019} are likely to have escaped from the density-bounded region that we have identified in this galaxy.
    
    \item Our analyses show that that dust porosity or dust covering fraction is an important factor in limiting the escape of Ly$\alpha$ photons.
    
    \item Using a sample of confirmed low-redshift LyC leakers, including SHOC 579, we found a moderate correlation between [SII]-deficiency and inferred $f^{esc}_{LyC}$. This correlation may become stronger in the future, when a large sample of confirmed LyC leakers is used with [SII]-deficiency having a large dynamic range. 
    
    \item We also found that the correlation mentioned above has a dependency on the gas-phase metallicity of LyC leakers. This suggests that metal-poor leakers are seen with a relatively high [SII]-deficiency and $f^{esc}_{LyC}$, regardless of their stellar masses. 
 
\end{itemize}}

 
\appendix
 
\section{[SIII] $\lambda$9069,9532 emission line fitting}\label{SIII-emana} 
 
\begin{figure}
\begin{center}
\rotatebox{0}{\includegraphics[width=0.48\textwidth]{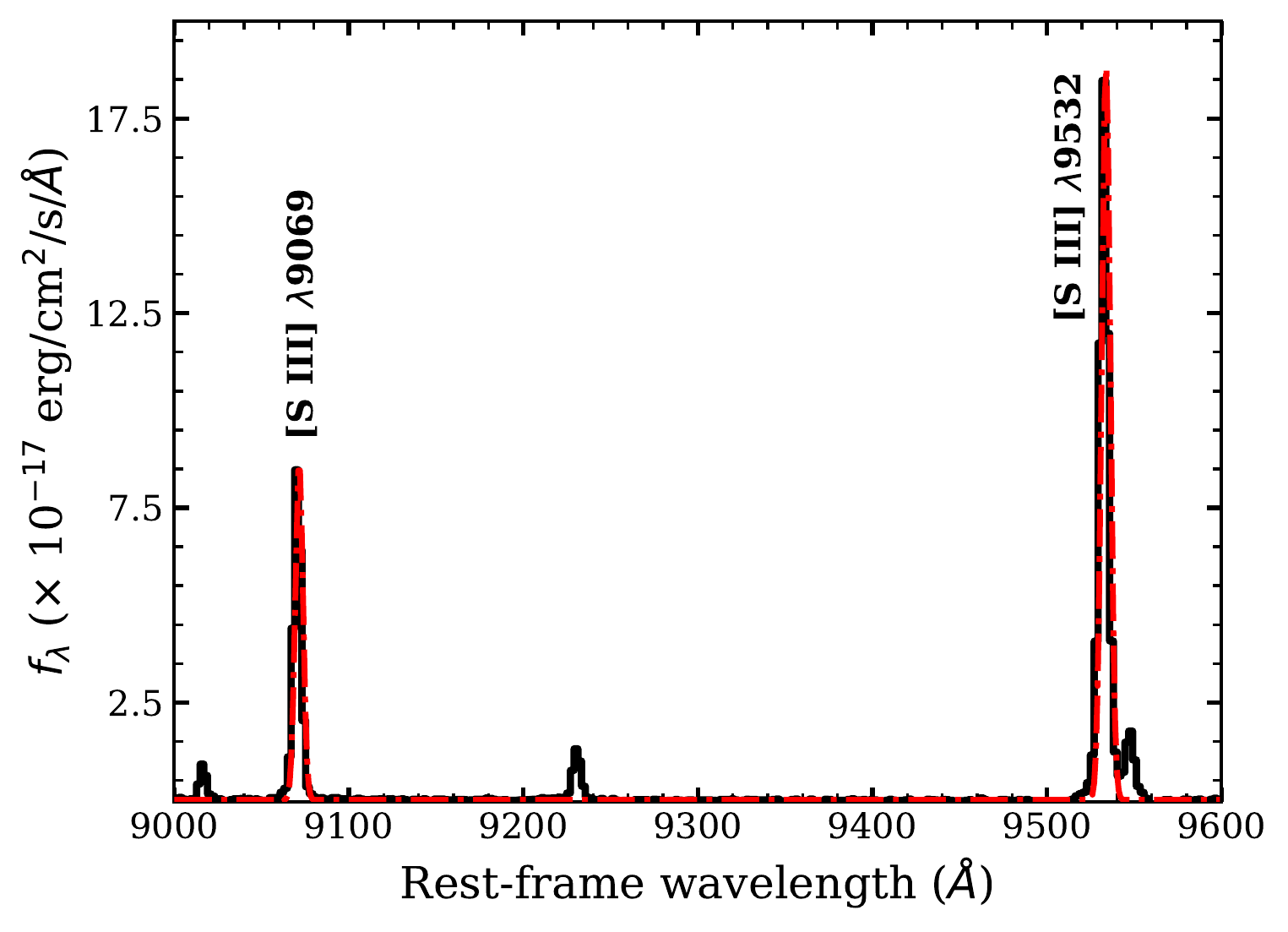}}
\rotatebox{0}{\includegraphics[width=0.44\textwidth] {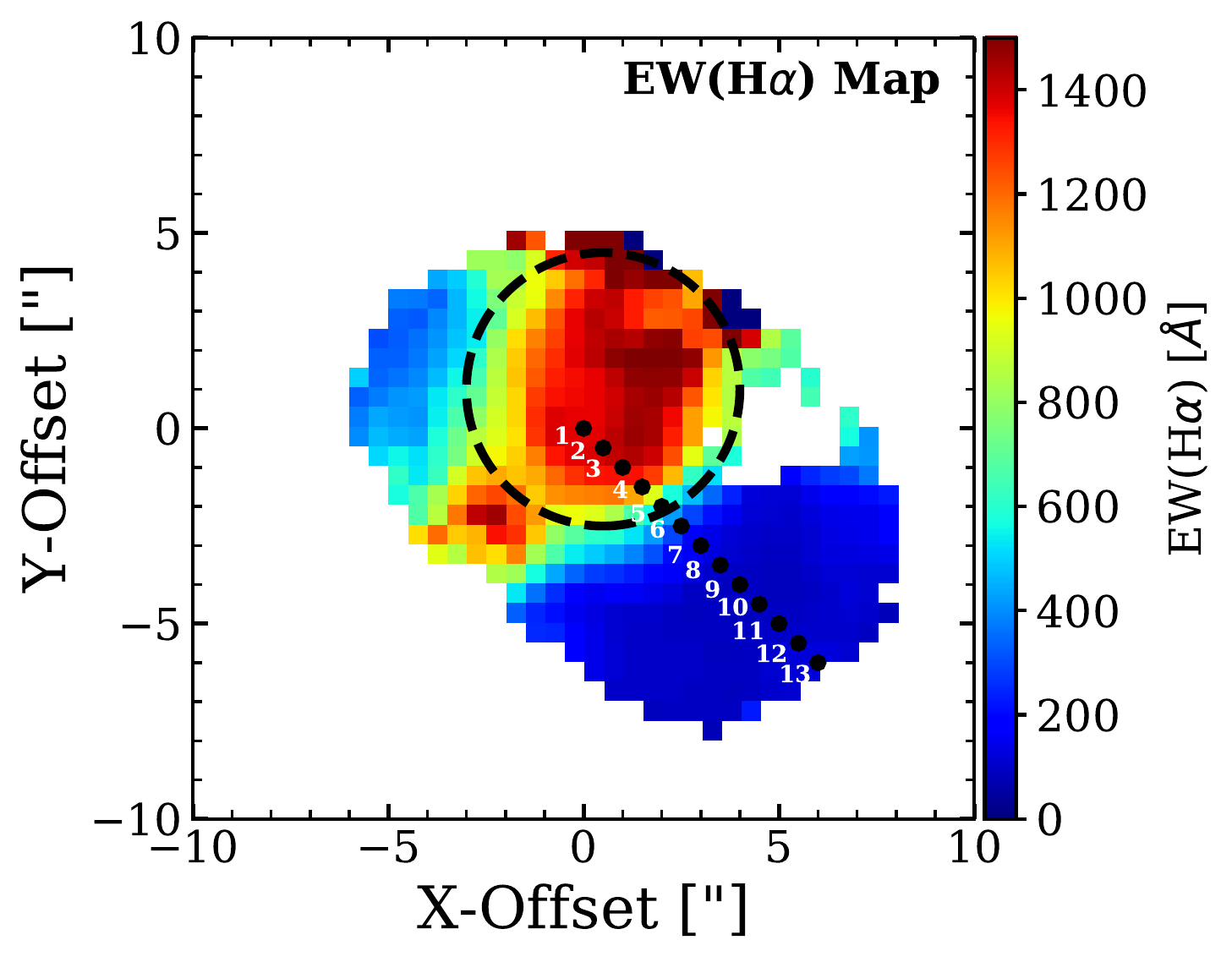}}
\rotatebox{0}{\includegraphics[width=0.45\textwidth]{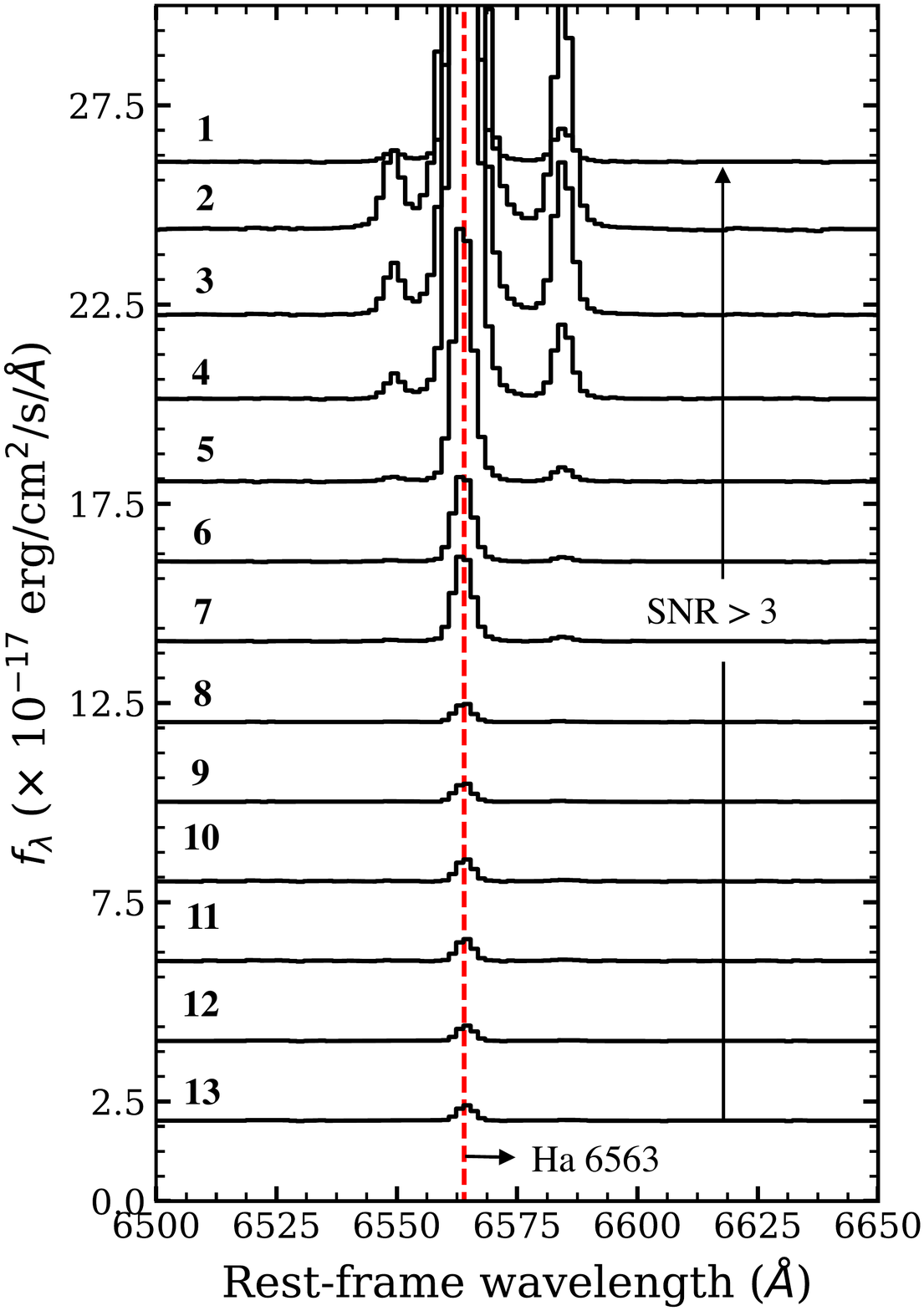}}
\rotatebox{0}{\includegraphics[width=0.47\textwidth,trim=0cm 0.5cm 0cm 0cm, clip=true]{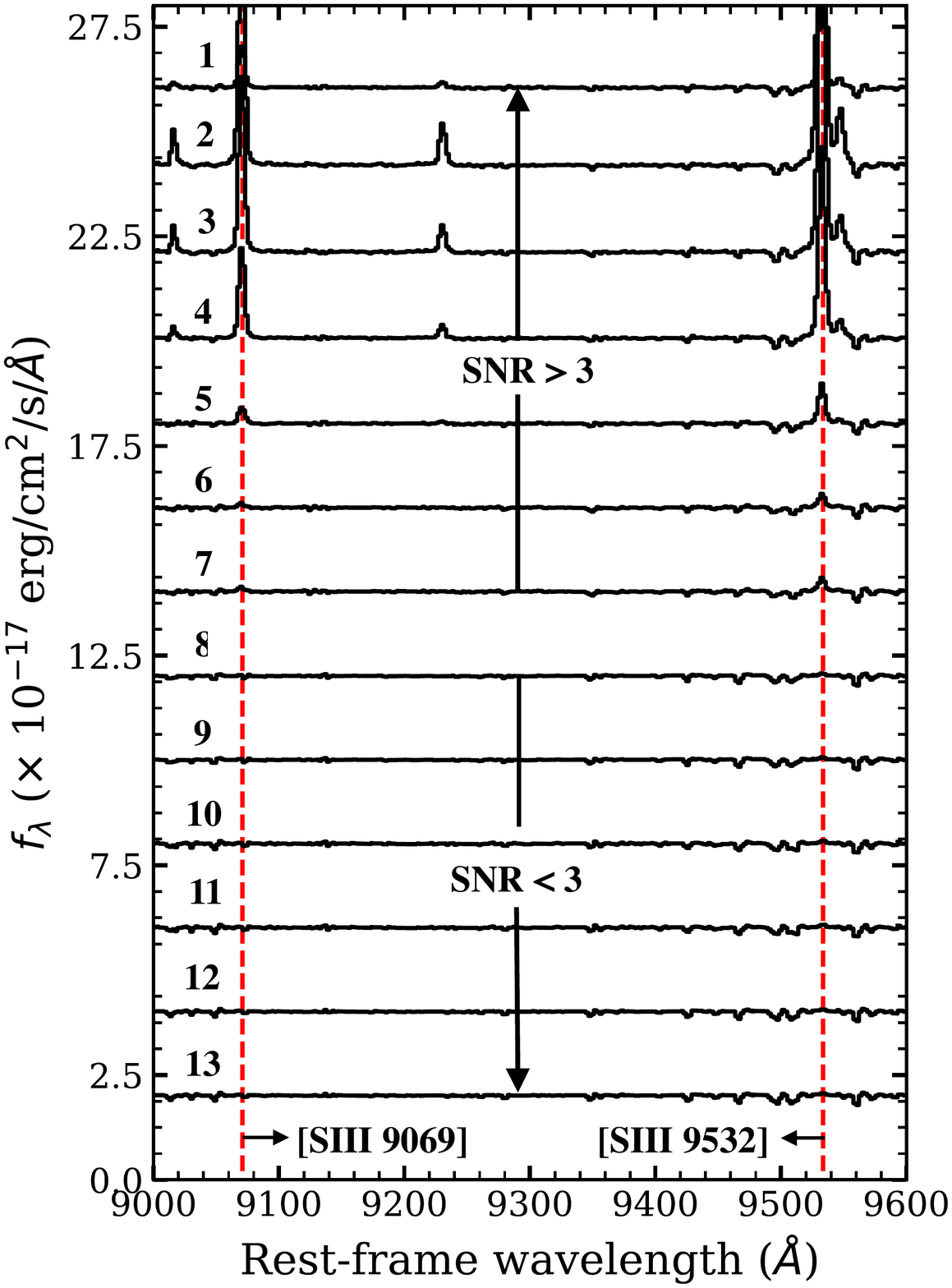}}
\caption{An example of the detection of two [SIII] $\lambda$9069 and [SIII] $\lambda$9532 emission lines from the density-bounded region of the galaxy (top left panel). The red dashed line indicates our fitted Gaussian models, after stellar continuum subtraction. The top right panel shows the H$\alpha$ EW map, where the drawn circle and dots ($1 - 13$) represent the density-bounded starburst region and radial spaxels from center to outer region of the galaxy, respectively. The bottom left and right panels show the detection of H$\alpha$ and [SIII] $\lambda$9069,9532 emission lines in the spectra along the radial direction of the galaxy as marked by spaxels $1 - 13$ shown in the top right panel.}
\label{SIII-fit}
\end{center}
\end{figure}
 
This section explains the estimation of [SIII] $\lambda$9069,9532 emission line flux used in the present study. Since the MILES stellar library fits only optical spectrum in the wavelength range of $\sim 3900 - 7400$ \AA, we therefore performed our own Gaussian model fit to the [SIII] $\lambda$9069,9532 emission line after subtracting the local stellar continuum flux estimated from the line-free region (continuum) on both sides of the emission lines. Thereafter, the emission line flux is directly measured using area under the fitted Gaussian models. In our fitting process, we considered only those galaxy spaxels whose [SIII] $\lambda$9069,9532 emission line has good detection (i.e., SNR $\geqslant$ 3). For example, Fig.~\ref{SIII-fit} (top left panel) shows our typical fit to the [SIII] $\lambda$9069,9532 emission line at central (X = 37, Y = 37) spaxel of the galaxy. This figure also indicates that the wavelength region around [SIII] $\lambda$9069,9532 emission line is free from the sky-lines contamination. We notice that a clear detection of [SIII] $\lambda$9069,9532 emission line is mainly made around starburst region marked by the circle in H$\alpha$ EW-map (see top right panel of Fig.~\ref{SIII-fit}), while it is not detected due to low SNR (i.e., SNR $\textless$ 3; except a few spaxels) in the extended disk region i.e., outside the circle. In Fig.~\ref{SIII-fit}, this argument is well-explained by displaying a set of spectra of H$\alpha$ (bottom left) and [SIII] $\lambda$9069,9532 (bottom right) emission lines along the radial direction from the center to outer spaxels (marked in the top right panel) of the galaxy. Due to non-detection of [SIII] $\lambda$9069,9532 emission line, it could not be possible to construct the map of S$_{32}$ parameter (see Fig.~\ref{SIII-ana}) in the extended stellar disk of the galaxy.

\bibliography{ms-DB}

\end{document}